\documentclass[namedreferences]{solarphysics}

\usepackage[hyperref,optionalrh]{spr-sola-addons} 
\usepackage[all]{hypcap} 
\usepackage{graphicx}        
\usepackage{amssymb}        
\usepackage{textcmds}        
\usepackage{color}           
\usepackage{breakurl}        

\hypersetup{
	colorlinks = true, 
	allcolors = blue,
}

\makeatletter
\newcommand\tsb[1]{\@textsubscript{\selectfont#1}}
\def\@textsubscript#1{{\m@th\ensuremath{_{\mbox{\fontsize\sf@size\z@#1}}}}}
\newcommand\tsp[1]{\@textsuperscript{\selectfont#1}}
\def\@textsuperscript#1{{\m@th\ensuremath{^{\mbox{\fontsize\sf@size\z@#1}}}}}

\providecommand{\e}[1]{\ensuremath{\times 10^{#1}}}
\providecommand{\lsr}[1]{\textit{V}\tsb{LSR}}

\providecommand{\rsolar}[1]{#1 R\tsb{$\odot$}}
\providecommand{\kms}{km s\tsp{-1}}

\providecommand{\msq}{m s\tsp{-2}}

\usepackage{multirow}

 \renewcommand*{\thetabnote}{\alph{tabnote}}

 \providecommand{\TypeARg}{135} \providecommand{\TypeIPg}{76} \providecommand{\TypeQSg}{392}
\providecommand{\TypeQSg}{392} \providecommand{\TypePCg}{122} \providecommand{\TypeTEg}{4} \providecommand{\TypeQg}{20}
\providecommand{\TypeQg}{20}  \providecommand{\TypeARq}{37} \providecommand{\TypeIPq}{48}
\providecommand{\TypeIPq}{48} \providecommand{\TypeQSq}{59} \providecommand{\TypePCq}{10} \providecommand{\TypeTEq}{1}
\providecommand{\TypeTEq}{1}  \providecommand{\TypeAll}{904} \providecommand{\TypeAR}{172}
\providecommand{\TypeAR}{172} \providecommand{\TypeIP}{124} \providecommand{\TypeQS}{451} \providecommand{\TypePC}{132}
\providecommand{\TypePC}{132} \providecommand{\TypeTE}{5} \providecommand{\TypeQ}{20}

 \providecommand{\AllPosDg}{401} \providecommand{\ARPosDg}{92} \providecommand{\IPPosDg}{68}
\providecommand{\IPPosDg}{68} \providecommand{\QSPosDg}{214} \providecommand{\PCPosDg}{19} \providecommand{\TEPosDg}{5}
\providecommand{\TEPosDg}{5} \providecommand{\QPosDg}{3}  
   
   \providecommand{\AllPosD}{401}
\providecommand{\AllPosD}{401}

  \providecommand{\AllPosLg}{363} \providecommand{\ARPosLg}{50}
\providecommand{\ARPosLg}{50} \providecommand{\IPPosLg}{48} \providecommand{\QSPosLg}{178} \providecommand{\PCPosLg}{82}
\providecommand{\PCPosLg}{82} \providecommand{\TEPosLg}{0} \providecommand{\QPosLg}{5}

 \providecommand{\AllPosL}{363}

   \providecommand{\AllPosBg}{139}
\providecommand{\AllPosBg}{139} \providecommand{\ARPosBg}{30} \providecommand{\IPPosBg}{8} \providecommand{\QSPosBg}{58}
\providecommand{\QSPosBg}{58} \providecommand{\PCPosBg}{31} \providecommand{\TEPosBg}{0} \providecommand{\QPosBg}{12}
\providecommand{\QPosBg}{12}   
   
  \providecommand{\AllPosB}{139}

 \providecommand{\AllEruFg}{285} \providecommand{\AREruFg}{56} \providecommand{\IPEruFg}{39}
\providecommand{\IPEruFg}{39} \providecommand{\QSEruFg}{143} \providecommand{\PCEruFg}{36} \providecommand{\TEEruFg}{3}
\providecommand{\TEEruFg}{3} \providecommand{\QEruFg}{8} \providecommand{\AllEruFq}{73} \providecommand{\AREruFq}{13}
\providecommand{\AREruFq}{13} \providecommand{\IPEruFq}{7} \providecommand{\QSEruFq}{42} \providecommand{\PCEruFq}{9}
\providecommand{\PCEruFq}{9}  \providecommand{\QEruFq}{2} \providecommand{\AllEruF}{358}
\providecommand{\AllEruF}{358}

  \providecommand{\AllEruPg}{266} \providecommand{\AREruPg}{38}
\providecommand{\AREruPg}{38} \providecommand{\IPEruPg}{38} \providecommand{\QSEruPg}{131} \providecommand{\PCEruPg}{53}
\providecommand{\PCEruPg}{53} \providecommand{\TEEruPg}{2} \providecommand{\QEruPg}{4} \providecommand{\AllEruPq}{99}
\providecommand{\AllEruPq}{99} \providecommand{\AREruPq}{23} \providecommand{\IPEruPq}{17} \providecommand{\QSEruPq}{49}
\providecommand{\QSEruPq}{49} \providecommand{\PCEruPq}{7}  \providecommand{\QEruPq}{3}
\providecommand{\QEruPq}{3} \providecommand{\AllEruP}{365}

   \providecommand{\AllEruCg}{122}
\providecommand{\AllEruCg}{122} \providecommand{\AREruCg}{33} \providecommand{\IPEruCg}{15} \providecommand{\QSEruCg}{57}
\providecommand{\QSEruCg}{57} \providecommand{\PCEruCg}{14} \providecommand{\TEEruCg}{0} \providecommand{\QEruCg}{3}
\providecommand{\QEruCg}{3} \providecommand{\AllEruCq}{56} \providecommand{\AREruCq}{9} \providecommand{\IPEruCq}{7}
\providecommand{\IPEruCq}{7} \providecommand{\QSEruCq}{29} \providecommand{\PCEruCq}{11} 
 \providecommand{\QEruCq}{0} \providecommand{\AllEruC}{178}

 \providecommand{\AllEruOg}{18} \providecommand{\AREruOg}{5} \providecommand{\IPEruOg}{1}
\providecommand{\IPEruOg}{1} \providecommand{\QSEruOg}{10} \providecommand{\PCEruOg}{2} \providecommand{\TEEruOg}{0}
\providecommand{\TEEruOg}{0} \providecommand{\QEruOg}{0} \providecommand{\AllEruOq}{2} \providecommand{\AREruOq}{1}
\providecommand{\AREruOq}{1}  \providecommand{\QSEruOq}{1} 
   \providecommand{\AllEruO}{20}
\providecommand{\AllEruO}{20}

  \providecommand{\AllSymSg}{307} \providecommand{\ARSymSg}{41}
\providecommand{\ARSymSg}{41} \providecommand{\IPSymSg}{49} \providecommand{\QSSymSg}{174} \providecommand{\PCSymSg}{34}
\providecommand{\PCSymSg}{34} \providecommand{\TESymSg}{2} \providecommand{\QSymSg}{7} \providecommand{\AllSymSq}{126}
\providecommand{\AllSymSq}{126} \providecommand{\ARSymSq}{30} \providecommand{\IPSymSq}{13} \providecommand{\QSSymSq}{62}
\providecommand{\QSSymSq}{62} \providecommand{\PCSymSq}{17} \providecommand{\TESymSq}{1} \providecommand{\QSymSq}{3}
\providecommand{\QSymSq}{3}   
   
  \providecommand{\AllSymSp}{48\%} \providecommand{\ARSymSp}{41\%}
\providecommand{\ARSymSp}{41\%} \providecommand{\IPSymSp}{50\%} \providecommand{\QSSymSp}{52\%} \providecommand{\PCSymSp}{39\%}
\providecommand{\PCSymSp}{39\%}   \providecommand{\AllSymAg}{232}
\providecommand{\AllSymAg}{232} \providecommand{\ARSymAg}{49} \providecommand{\IPSymAg}{29} \providecommand{\QSSymAg}{116}
\providecommand{\QSSymAg}{116} \providecommand{\PCSymAg}{33} \providecommand{\TESymAg}{1} \providecommand{\QSymAg}{4}
\providecommand{\QSymAg}{4} \providecommand{\AllSymAq}{114} \providecommand{\ARSymAq}{19} \providecommand{\IPSymAq}{16}
\providecommand{\IPSymAq}{16} \providecommand{\QSSymAq}{48} \providecommand{\PCSymAq}{27} \providecommand{\TESymAq}{1}
\providecommand{\TESymAq}{1} \providecommand{\QSymAq}{3}  
   
   \providecommand{\AllSymAp}{38\%}
\providecommand{\AllSymAp}{38\%} \providecommand{\ARSymAp}{40\%} \providecommand{\IPSymAp}{36\%} \providecommand{\QSSymAp}{36\%}
\providecommand{\QSSymAp}{36\%} \providecommand{\PCSymAp}{45\%}  
 \providecommand{\AllSymQg}{125} \providecommand{\ARSymQg}{33} \providecommand{\IPSymQg}{17}
\providecommand{\IPSymQg}{17} \providecommand{\QSSymQg}{51} \providecommand{\PCSymQg}{21} \providecommand{\TESymQg}{0}
\providecommand{\TESymQg}{0} \providecommand{\QSymQg}{3}

 \providecommand{\AllSymQp}{14\%}  
   
  \providecommand{\AllDirRg}{625} \providecommand{\ARDirRg}{95}
\providecommand{\ARDirRg}{95} \providecommand{\IPDirRg}{79} \providecommand{\QSDirRg}{338} \providecommand{\PCDirRg}{97}
\providecommand{\PCDirRg}{97} \providecommand{\TEDirRg}{2} \providecommand{\QDirRg}{14} \providecommand{\AllDirRq}{51}
\providecommand{\AllDirRq}{51} \providecommand{\ARDirRq}{12} \providecommand{\IPDirRq}{3} \providecommand{\QSDirRq}{26}
\providecommand{\QSDirRq}{26} \providecommand{\PCDirRq}{7} \providecommand{\TEDirRq}{2} \providecommand{\QDirRq}{1}
\providecommand{\QDirRq}{1}   
   
  \providecommand{\AllDirRp}{75\%} 
   
   \providecommand{\AllDirNRg}{73}
\providecommand{\AllDirNRg}{73} \providecommand{\ARDirNRg}{11} \providecommand{\IPDirNRg}{12} \providecommand{\QSDirNRg}{39}
\providecommand{\QSDirNRg}{39} \providecommand{\PCDirNRg}{10} \providecommand{\TEDirNRg}{0} \providecommand{\QDirNRg}{1}
\providecommand{\QDirNRg}{1} \providecommand{\AllDirNRq}{35} \providecommand{\ARDirNRq}{6} \providecommand{\IPDirNRq}{4}
\providecommand{\IPDirNRq}{4} \providecommand{\QSDirNRq}{17} \providecommand{\PCDirNRq}{6} 
 \providecommand{\QDirNRq}{2}  
   
   \providecommand{\AllDirNRp}{12\%}
\providecommand{\AllDirNRp}{12\%} \providecommand{\ARDirNRp}{10\%} \providecommand{\IPDirNRp}{13\%} \providecommand{\QSDirNRp}{12\%}
\providecommand{\QSDirNRp}{12\%} \providecommand{\PCDirNRp}{12\%}  
 \providecommand{\AllDirSg}{72} \providecommand{\ARDirSg}{36} \providecommand{\IPDirSg}{18}
\providecommand{\IPDirSg}{18} \providecommand{\QSDirSg}{12} \providecommand{\PCDirSg}{5} \providecommand{\TEDirSg}{1}
\providecommand{\TEDirSg}{1} \providecommand{\QDirSg}{0} \providecommand{\AllDirSq}{26} \providecommand{\ARDirSq}{8}
\providecommand{\ARDirSq}{8} \providecommand{\IPDirSq}{5} \providecommand{\QSDirSq}{8} \providecommand{\PCDirSq}{4}
\providecommand{\PCDirSq}{4}  \providecommand{\QDirSq}{1}

 \providecommand{\AllDirSp}{11\%} \providecommand{\ARDirSp}{26\%} \providecommand{\IPDirSp}{19\%}
\providecommand{\IPDirSp}{19\%} \providecommand{\QSDirSp}{4\%} \providecommand{\PCDirSp}{7\%} 
  \providecommand{\AllDirQg}{22} \providecommand{\ARDirQg}{4}
\providecommand{\ARDirQg}{4} \providecommand{\IPDirQg}{3} \providecommand{\QSDirQg}{11} \providecommand{\PCDirQg}{3}
\providecommand{\PCDirQg}{3} \providecommand{\TEDirQg}{0} \providecommand{\QDirQg}{1}

  \providecommand{\AllDirQp}{2\%} 
   
   \providecommand{\AllTwiYg}{408}
\providecommand{\AllTwiYg}{408} \providecommand{\ARTwiYg}{70} \providecommand{\IPTwiYg}{65} \providecommand{\QSTwiYg}{201}
\providecommand{\QSTwiYg}{201} \providecommand{\PCTwiYg}{58} \providecommand{\TETwiYg}{1} \providecommand{\QTwiYg}{13}
\providecommand{\QTwiYg}{13} \providecommand{\AllTwiYq}{130} \providecommand{\ARTwiYq}{31} \providecommand{\IPTwiYq}{19}
\providecommand{\IPTwiYq}{19} \providecommand{\QSTwiYq}{63} \providecommand{\PCTwiYq}{14} 
 \providecommand{\QTwiYq}{3}  
   
   \providecommand{\AllTwiYp}{60\%}
\providecommand{\AllTwiYp}{60\%} \providecommand{\ARTwiYp}{59\%} \providecommand{\IPTwiYp}{68\%} \providecommand{\QSTwiYp}{59\%}
\providecommand{\QSTwiYp}{59\%} \providecommand{\PCTwiYp}{55\%}  
 \providecommand{\AllTwiNg}{265} \providecommand{\ARTwiNg}{44} \providecommand{\IPTwiNg}{28}
\providecommand{\IPTwiNg}{28} \providecommand{\QSTwiNg}{143} \providecommand{\PCTwiNg}{47} \providecommand{\TETwiNg}{1}
\providecommand{\TETwiNg}{1} \providecommand{\QTwiNg}{2} \providecommand{\AllTwiNq}{68} \providecommand{\ARTwiNq}{14}
\providecommand{\ARTwiNq}{14} \providecommand{\IPTwiNq}{8} \providecommand{\QSTwiNq}{34} \providecommand{\PCTwiNq}{10}
\providecommand{\PCTwiNq}{10} \providecommand{\TETwiNq}{1} \providecommand{\QTwiNq}{1}

 \providecommand{\AllTwiNp}{37\%}  
   
  \providecommand{\AllTwiQg}{32} \providecommand{\ARTwiQg}{13}
\providecommand{\ARTwiQg}{13} \providecommand{\IPTwiQg}{4} \providecommand{\QSTwiQg}{9} \providecommand{\PCTwiQg}{3}
\providecommand{\PCTwiQg}{3} \providecommand{\TETwiQg}{2} \providecommand{\QTwiQg}{1}

   \providecommand{\AllKinYg}{73}
\providecommand{\AllKinYg}{73} \providecommand{\ARKinYg}{26} \providecommand{\IPKinYg}{15} \providecommand{\QSKinYg}{27}
\providecommand{\QSKinYg}{27} \providecommand{\PCKinYg}{1} \providecommand{\TEKinYg}{0} \providecommand{\QKinYg}{4}
\providecommand{\QKinYg}{4} \providecommand{\AllKinYq}{71} \providecommand{\ARKinYq}{14} \providecommand{\IPKinYq}{10}
\providecommand{\IPKinYq}{10} \providecommand{\QSKinYq}{38} \providecommand{\PCKinYq}{7} 
 \providecommand{\QKinYq}{2} \providecommand{\AllKinY}{144} 
   
   \providecommand{\AllKinYp}{16\%}
\providecommand{\AllKinYp}{16\%} \providecommand{\ARKinYp}{23\%} \providecommand{\IPKinYp}{20\%} \providecommand{\QSKinYp}{14\%}
\providecommand{\QSKinYp}{14\%} \providecommand{\PCKinYp}{6\%}  
 \providecommand{\AllKinNg}{656} \providecommand{\ARKinNg}{102} \providecommand{\IPKinNg}{79}
\providecommand{\IPKinNg}{79} \providecommand{\QSKinNg}{350} \providecommand{\PCKinNg}{110} \providecommand{\TEKinNg}{5}
\providecommand{\TEKinNg}{5} \providecommand{\QKinNg}{10} \providecommand{\AllKinNq}{61} \providecommand{\ARKinNq}{14}
\providecommand{\ARKinNq}{14} \providecommand{\IPKinNq}{13} \providecommand{\QSKinNq}{23} \providecommand{\PCKinNq}{10}
\providecommand{\PCKinNq}{10}  \providecommand{\QKinNq}{1}

 \providecommand{\AllKinNp}{79\%}  
   
  \providecommand{\AllKinQg}{43} \providecommand{\ARKinQg}{16}
\providecommand{\ARKinQg}{16} \providecommand{\IPKinQg}{7} \providecommand{\QSKinQg}{13} \providecommand{\PCKinQg}{4}
\providecommand{\PCKinQg}{4} \providecommand{\TEKinQg}{0} \providecommand{\QKinQg}{3}

  \providecommand{\AllKinQp}{5\%} 
   
   \providecommand{\AllVerYg}{384}
\providecommand{\AllVerYg}{384} \providecommand{\ARVerYg}{10} \providecommand{\IPVerYg}{52} \providecommand{\QSVerYg}{246}
\providecommand{\QSVerYg}{246} \providecommand{\PCVerYg}{69} \providecommand{\TEVerYg}{4} \providecommand{\QVerYg}{3}
\providecommand{\QVerYg}{3} \providecommand{\AllVerYq}{58} \providecommand{\ARVerYq}{6} \providecommand{\IPVerYq}{16}
\providecommand{\IPVerYq}{16} \providecommand{\QSVerYq}{30} \providecommand{\PCVerYq}{6} 
   \providecommand{\ARVerY}{16}
\providecommand{\ARVerY}{16}   
   \providecommand{\AllVerYp}{58\%}
\providecommand{\AllVerYp}{58\%} \providecommand{\ARVerYp}{11\%} \providecommand{\IPVerYp}{59\%} \providecommand{\QSVerYp}{70\%}
\providecommand{\QSVerYp}{70\%} \providecommand{\PCVerYp}{74\%}  
 \providecommand{\AllVerNg}{296} \providecommand{\ARVerNg}{118} \providecommand{\IPVerNg}{42}
\providecommand{\IPVerNg}{42} \providecommand{\QSVerNg}{105} \providecommand{\PCVerNg}{26} \providecommand{\TEVerNg}{0}
\providecommand{\TEVerNg}{0} \providecommand{\QVerNg}{5} \providecommand{\AllVerNq}{26} \providecommand{\ARVerNq}{8}
\providecommand{\ARVerNq}{8} \providecommand{\IPVerNq}{6} \providecommand{\QSVerNq}{11} 
 \providecommand{\TEVerNq}{1}

 \providecommand{\AllVerNp}{42\%}  
   
  \providecommand{\AllCavYg}{97} \providecommand{\ARCavYg}{2}
\providecommand{\ARCavYg}{2} \providecommand{\IPCavYg}{10} \providecommand{\QSCavYg}{47} \providecommand{\PCCavYg}{37}
\providecommand{\PCCavYg}{37} \providecommand{\TECavYg}{0} \providecommand{\QCavYg}{1} \providecommand{\AllCavYq}{27}
\providecommand{\AllCavYq}{27} \providecommand{\ARCavYq}{2}  \providecommand{\QSCavYq}{12}
\providecommand{\QSCavYq}{12} \providecommand{\PCCavYq}{12}  \providecommand{\QCavYq}{0}
\providecommand{\QCavYq}{0}   
   
  \providecommand{\AllCavYp}{34\%} \providecommand{\ARCavYp}{8\%}
\providecommand{\ARCavYp}{8\%} \providecommand{\IPCavYp}{23\%} \providecommand{\QSCavYp}{33\%} \providecommand{\PCCavYp}{60\%}
\providecommand{\PCCavYp}{60\%}   \providecommand{\AllCavNg}{165}
\providecommand{\AllCavNg}{165} \providecommand{\ARCavNg}{35} \providecommand{\IPCavNg}{25} \providecommand{\QSCavNg}{79}
\providecommand{\QSCavNg}{79} \providecommand{\PCCavNg}{25} \providecommand{\TECavNg}{0} \providecommand{\QCavNg}{1}
\providecommand{\QCavNg}{1} \providecommand{\AllCavNq}{13}  
 \providecommand{\QSCavNq}{8} \providecommand{\PCCavNq}{5}

   \providecommand{\AllCavNp}{49\%}
\providecommand{\AllCavNp}{49\%}   
   
 \providecommand{\AllCavDg}{47} \providecommand{\ARCavDg}{9} \providecommand{\IPCavDg}{9}
\providecommand{\IPCavDg}{9} \providecommand{\QSCavDg}{24} \providecommand{\PCCavDg}{3} \providecommand{\TECavDg}{0}
\providecommand{\TECavDg}{0} \providecommand{\QCavDg}{2} \providecommand{\AllCavDq}{14} \providecommand{\ARCavDq}{2}
\providecommand{\ARCavDq}{2} \providecommand{\IPCavDq}{3} \providecommand{\QSCavDq}{8} 
  \providecommand{\QCavDq}{1}

 \providecommand{\AllCavDp}{17\%} \providecommand{\ARCavDp}{22\%} \providecommand{\IPCavDp}{25\%}
\providecommand{\IPCavDp}{25\%} \providecommand{\QSCavDp}{18\%} \providecommand{\PCCavDp}{4\%} 
  \providecommand{\AllCMEYg}{609} \providecommand{\ARCMEYg}{108}
\providecommand{\ARCMEYg}{108} \providecommand{\IPCMEYg}{85} \providecommand{\QSCMEYg}{307} \providecommand{\PCCMEYg}{91}
\providecommand{\PCCMEYg}{91} \providecommand{\TECMEYg}{4} \providecommand{\QCMEYg}{14} \providecommand{\AllCMEYq}{44}
\providecommand{\AllCMEYq}{44} \providecommand{\ARCMEYq}{12} \providecommand{\IPCMEYq}{8} \providecommand{\QSCMEYq}{18}
\providecommand{\QSCMEYq}{18} \providecommand{\PCCMEYq}{5}  \providecommand{\QCMEYq}{1}
\providecommand{\QCMEYq}{1}   
   
  \providecommand{\AllCMEYp}{72\%} \providecommand{\ARCMEYp}{70\%}
\providecommand{\ARCMEYp}{70\%} \providecommand{\IPCMEYp}{75\%} \providecommand{\QSCMEYp}{72\%} \providecommand{\PCCMEYp}{73\%}
\providecommand{\PCCMEYp}{73\%}   \providecommand{\AllCMENg}{225}
\providecommand{\AllCMENg}{225} \providecommand{\ARCMENg}{50} \providecommand{\IPCMENg}{26} \providecommand{\QSCMENg}{112}
\providecommand{\QSCMENg}{112} \providecommand{\PCCMENg}{32} \providecommand{\TECMENg}{1} \providecommand{\QCMENg}{4}
\providecommand{\QCMENg}{4} \providecommand{\AllCMENq}{26} \providecommand{\ARCMENq}{2} \providecommand{\IPCMENq}{5}
\providecommand{\IPCMENq}{5} \providecommand{\QSCMENq}{14} \providecommand{\PCCMENq}{4} 
 \providecommand{\QCMENq}{1}  
   
   \providecommand{\AllCMENp}{28\%}
\providecommand{\AllCMENp}{28\%}   
   
 \providecommand{\AllFlaYg}{222} \providecommand{\ARFlaYg}{94} \providecommand{\IPFlaYg}{50}
\providecommand{\IPFlaYg}{50} \providecommand{\QSFlaYg}{68} \providecommand{\PCFlaYg}{7} \providecommand{\TEFlaYg}{1}
\providecommand{\TEFlaYg}{1} \providecommand{\QFlaYg}{2} \providecommand{\AllFlaYq}{15} \providecommand{\ARFlaYq}{2}
\providecommand{\ARFlaYq}{2} \providecommand{\IPFlaYq}{3} \providecommand{\QSFlaYq}{9} \providecommand{\PCFlaYq}{1}
\providecommand{\PCFlaYq}{1}

  \providecommand{\AllFlaNg}{682} \providecommand{\ARFlaNg}{78}
\providecommand{\ARFlaNg}{78} \providecommand{\IPFlaNg}{74} \providecommand{\QSFlaNg}{383} \providecommand{\PCFlaNg}{125}
\providecommand{\PCFlaNg}{125} \providecommand{\TEFlaNg}{4} \providecommand{\QFlaNg}{18} \providecommand{\AllFlaNq}{15}
\providecommand{\AllFlaNq}{15} \providecommand{\ARFlaNq}{2} \providecommand{\IPFlaNq}{3} \providecommand{\QSFlaNq}{9}
\providecommand{\QSFlaNq}{9} \providecommand{\PCFlaNq}{1}

\begin{document}

\begin{article}

\begin{opening}

\title{Prominence and Filament Eruptions Observed by the Solar Dynamics Observatory: 
Statistical Properties, Kinematics, and Online Catalog}

\author[addressref={af1},corref,email={pmccauley@cfa.harvard.edu}]{\inits{P.~I.}\fnm{P.~I.}~\lnm{McCauley}}
\author[addressref={af2,af1},corref,email={ynsu@pmo.ac.cn}]{\inits{Y.}\fnm{Y.}~\lnm{Su}}
\author[addressref={af1}]{\inits{N.}\fnm{N.}~\lnm{Schanche}}
\author[addressref={af3,af1}]{\inits{K.~E.}\fnm{K.~E.}~\lnm{Evans}}
\author[addressref={af4,af2}]{\inits{C.}\fnm{C.}~\lnm{Su}}
\author[addressref={af1}]{\inits{S.}\fnm{S.}~\lnm{McKillop}}
\author[addressref={af1}]{\inits{K.~K.}\fnm{K.~K.}~\lnm{Reeves}}

\address[id=af1]{Harvard-Smithsonian Center for Astrophysics, Cambridge, MA 02138, USA}
\address[id=af2]{Key Laboratory for Dark Matter and Space Science, Purple Mountain Observatory, Chinese Academy of Sciences, Nanjing 210008, China}
\address[id=af3]{Department of Astronomy, University of Maryland, College Park, MD 20742, USA}
\address[id=af4]{School of Astronomy and Space Science, Nanjing University, Nanjing 210093, China}



\runningauthor{McCauley et al.}
\runningtitle{Prominence Eruptions Observed by the SDO}


\begin{abstract}

We present a statistical study of prominence and filament eruptions observed by the Atmospheric Imaging Assembly (AIA) aboard 
the Solar Dynamics Observatory (SDO). 
Several properties are recorded for \TypeAll{} events that were culled from the 
Heliophysics Event Knowledgebase (HEK) and incorporated into an online catalog for general use. 
These characteristics include the filament and eruption type, eruption symmetry and direction, 
apparent twisting and writhing motions, and the presence of vertical threads and coronal cavities. 
Associated flares and white-light coronal mass ejections (CME) are also recorded. 
Total rates are given for each property along with how they differ among filament types.
We also examine the 
kinematics of 106 limb events to characterize the distinct slow- and fast-rise phases often exhibited by filament eruptions. 
The average fast-rise onset height, slow-rise duration, 
slow-rise velocity, maximum field-of-view (FOV) velocity, and maximum FOV acceleration are 83 Mm, 4.4 hours, 2.1 \kms{}, 
106 \kms{}, and 111 \msq{}, respectively. All parameters exhibit lognormal probability distributions similar 
to that of CME speeds. 
A positive correlation between latitude and fast-rise onset height is found, 
which we attribute to a corresponding negative correlation in the 
average vertical magnetic field gradient, or decay index, estimated 
from potential field source surface (PFSS) extrapolations. 
We also find the decay index at the fast-rise onset point to be 1.1 on 
average, consistent with the critical instability threshold theorized for straight current channels.  
Finally, we explore relationships between the derived kinematics properties 
and apparent twisting motions. 
We find that events with evident twist have significantly faster CME speeds 
and significantly lower fast-rise onset heights, suggesting relationships between 
these values and flux rope helicity.

\end{abstract}
\end{opening}


\section{Introduction} %
\label{introduction} %

Prominences are among the most common and well-studied features 
of the solar atmosphere, having been 
chronicled since the Middle Ages and identified as ``cloud formations" in the 
corona in the 1850s \citep{Tandberg98}. 
They are now recognized to be thin channels of relatively cool, dense plasma suspended in the corona 
by highly sheared or twisted magnetic fields above polarity inversion lines 
in the photospheric magnetic flux distribution.
Bright in H$\alpha{}$ emission when protruding from the limb, prominences 
appear dark compared to the surrounding chromosphere when seen on the disk, 
where they are referred to as filaments. 
We will use the terms \textit{prominence} and \textit{filament} interchangeably. 
Readers seeking detailed background information may direct their attention to the recent book, 
\textit{Solar Prominences} \citep{Vial15}, from which we reference several chapters, and the 
excellent review of prominence observations by \citet{Parenti14}. 


This paper focuses on prominence eruptions observed by the 
Atmospheric Imaging Assembly (AIA, \citealt{Lemen12}) aboard the 
Solar Dynamics Observatory (SDO, \citealt{Pesnell12}), launched in 2010.
The AIA provides continuous, full-sun coverage in 7 extreme ultraviolet (EUV) 
and 3 UV channels with high time (12s) and spatial (0.6$''$) resolution. 
A number of case studies have been published on AIA observations 
of eruptive prominences. 
\citet{Sterling11}, \citet{Tripathi13}, and \citet{Chen14} present 
separate observations of active region filament eruptions, each found to 
be consistent with an erupting flux-rope model. 
\citet{Su12,Su13} present observations 
and flux-rope modeling of a polar crown eruption, and \citet{Thompson13} 
further examines signs of twist in the same event. 
\citet{Williams13} estimate the mass of erupted filament material 
in a particularly stunning event, and several authors have combined 
the SDO with the Solar Terrestrial Relations Observatory 
(STEREO, \citealt{Howard08}) spacecraft to examine prominence eruptions from three separate viewing angles 
(e.g. \citealt{Chifu12, Koleva12}). 
\citet{Su12b} present a survey of 45 quiescent prominence eruptions observed by the 
SDO that serves as a pilot study for our work, with results generally consistent 
with those to be described here. 

We have recorded some basic observational properties for \TypeAll{} filament 
eruptions observed by the AIA, compiled an online catalog designed to 
aid the community in identifying promising events for future research, and 
performed a kinematics study using events selected from the catalog. 
\S\ref{observations} describes the catalog and the observations it provides. 
\S\ref{categories} details the recorded properties, including some background 
information on their significance and a report of our results. 
These include the filament type (\S\ref{type}), eruption symmetry (\S\ref{symmetry}) 
and direction (\S\ref{direction}), apparent twisting (\S\ref{twist}) and writhing (\S\ref{kink}) motions, the presence of 
vertical threads (\S\ref{threads}) and coronal cavities (\S\ref{cavity}), and white-light CME (\S\ref{cmes}) associations.
\S\ref{kinematics} details our kinematics study. This involves height-time measurements of 
106 limb eruptions, which are fit with an analytic approximation to provide statistics on the slow- 
and fast-rise phases often exhibited by filament eruptions. 
Our methods are detailed in \S\ref{procedure}, the analytic approximation is described in 
\S\ref{equation}, the impact of solar rotation is considered in \S\ref{rotation}, 
and our results are described in \S\ref{results}. 
\S\ref{discussion} discusses interesting patterns that emerge, including  
 a correlation between latitude and the fast-rise onset height (\S\ref{latitude}) along with suggestive relationships 
 between apparent twisting motions and the derived kinematics values (\S\ref{speed_twist}).
Finally, \S\ref{conclusion} summarizes our work.


 
\section{Online Catalog} %
\label{observations} %

 \begin{figure}
 \centerline{\includegraphics[width=\textwidth,clip=]{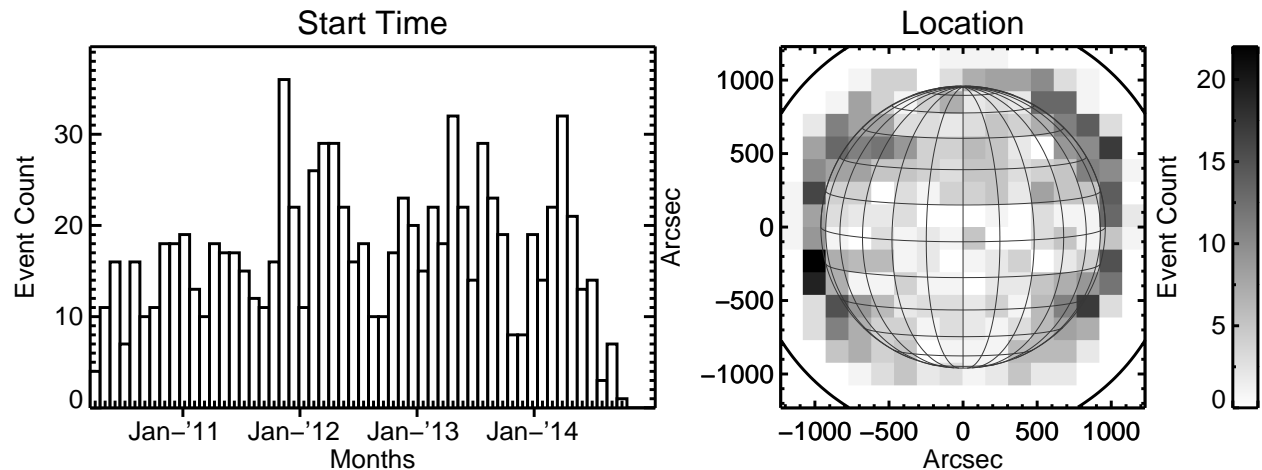}}
 \caption{Start time (left) and position (right) data for filament eruptions found in the HEK and 
 incorporated in the catalog.}
 \label{f-hek}
 \end{figure}

The catalog 
includes \TypeAll{} filament eruptions observed by the SDO between June of 2010 
and September of 2014. Times and positions are obtained from the 
Heliophysics Event Knowledgebase\footnote{Heliophysics Event Knowledgebase: \url{http://www.lmsal.com/hek/}} 
(\href{http://www.lmsal.com/hek/}{HEK}, \citealt{Hurlburt12}), which incorporates a wide variety 
of phenomena and relies on reports from individual observers for filament eruptions. 
Figure~\ref{f-hek} shows the event time and position distributions. An enhancement of eruptions near 
the limb is apparent, which is primarily due to the contribution of events behind the limb.
We also note the general position of each eruption, finding \AllPosD{} disk events, 
\AllPosL{} limb events, and \AllPosB{} events originating behind the limb. 

For each event, we provide cutout and full-disk movies in the 171, 193, and 304 \AA{} AIA passbands. 
If available, full-disk observations are also provided from 
the 304 \AA{} channels of the Extreme Ultraviolet Imagers (EUVI, \citealt{Wuelser04}) aboard the 
STEREO spacecraft. 
Images are obtained primarily from the 
\href{http://helioviewer.org/}{Helioviewer}\footnote{Helioviewer: \url{http://helioviewer.org/}} application program interface 
(\href{http://helioviewer.org/api/docs/v1/}{API}), which provides full-resolution JPEG2000 images 
that we then process separately to produce our content. 
In most cases, the 
171 \AA{} images are processed from the original 
level 1 data files using a radial filter described by \citet{Masson14}.
All movies are provided for download and may also be played via YouTube. 

Users may sort and filter the catalog based on the characteristics described in \S\ref{categories}. 
Additional comments on each event from 
the HEK and catalog observers are also provided and may be used to search the catalog for specific terms. 
Further details on the interface can be found at the 
\href{http://aia.cfa.harvard.edu/filament/}{catalog website}\footnote{SDO Filament Eruption Catalog: \url{http://aia.cfa.harvard.edu/filament/}}. 
Each event is also assigned an arbitrary rating based on how interesting a particular 
observer found it to be, which is intended to aid users who wish 
to browse through the large number of eruptions. A collection of movies from the highest-rated events is 
available as a YouTube playlist or as a downloadable TAR file, 
which may also be useful for education and public outreach. 
The catalog table may also be downloaded as an IDL save file or as a tab-delimited text file. 
Finally, we note that the structure of the website is essentially the 
same as that of the 
Hinode/XRT \href{http://xrt.cfa.harvard.edu/flare_catalog/}{Flare Catalog}\footnote{Hinode/XRT Flare Catalog: \url{http://xrt.cfa.harvard.edu/flare_catalog/}}  
and the 
Hinode \& SDO \href{http://aia.cfa.harvard.edu/sigmoid/}{Sigmoid Catalog}\footnote{Hinode \& SDO Sigmoid Catalog: \url{http://aia.cfa.harvard.edu/sigmoid/}} \citep{Savcheva14}.


\section{Statistical Properties}
\label{categories}

\begin{figure}
 \centerline{\includegraphics[width=.98\textwidth,clip=]{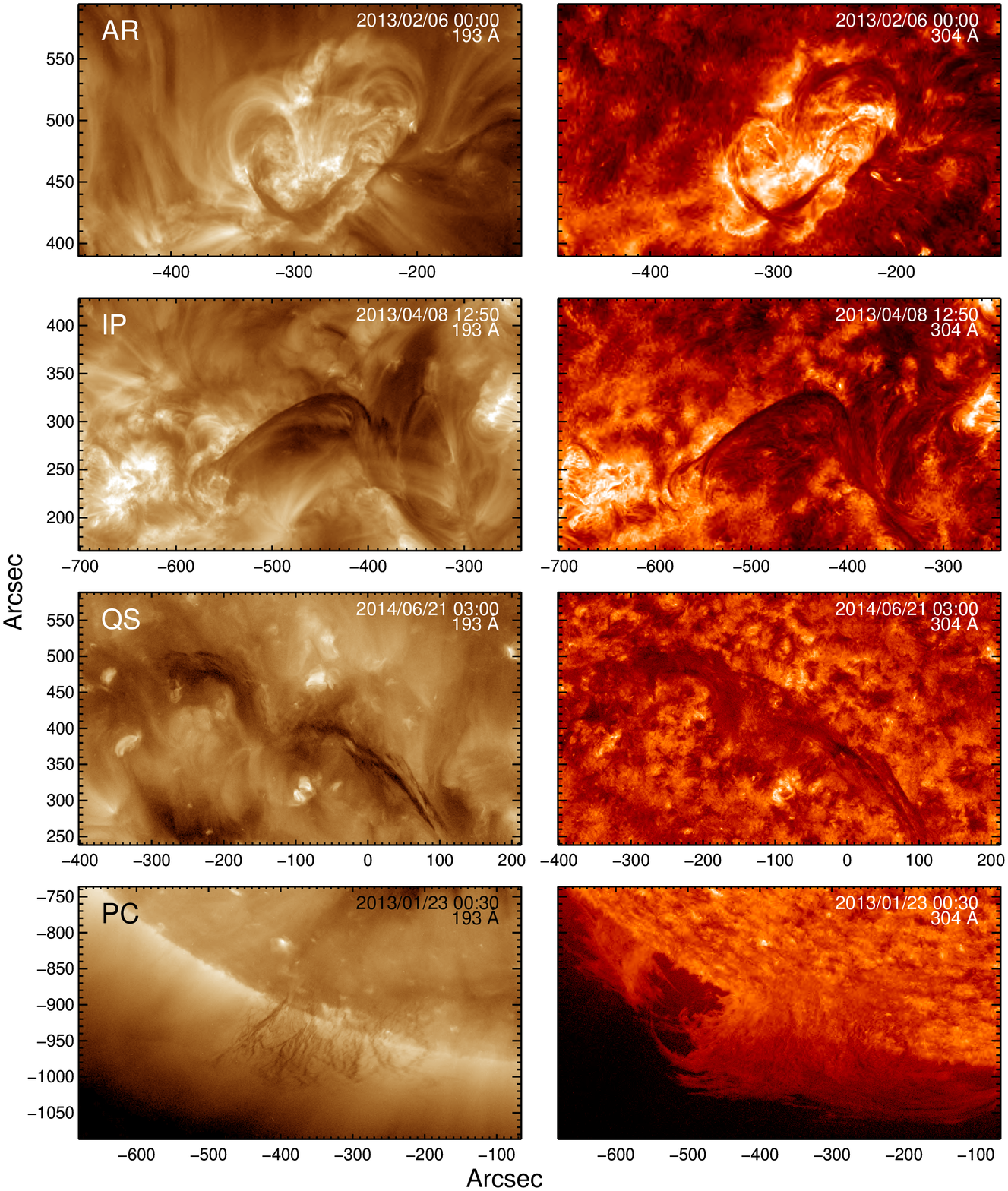}}
 \caption{Examples of the four main types: 
 active region (AR, \href{http://aia.cfa.harvard.edu/filament/index.html?search=0631}{No. 0631}), 
 intermediate (IP, \href{http://aia.cfa.harvard.edu/filament/index.html?search=0667}{No. 0667}), 
 quiescent (QS, \href{http://aia.cfa.harvard.edu/filament/index.html?search=0977}{No. 0977}), and 
 polar crown (PC, \href{http://aia.cfa.harvard.edu/filament/index.html?search=0621}{No. 0621}). 
 }
 \label{f-type}
 \end{figure}

The following subsections detail the properties recorded for each event, including  
brief background information on their significance and a report of our results. 
Each determination was made through manual inspection by one of three observers  
with the exception of flare associations, which are automatically populated using 
data from the ``\href{http://www.lmsal.com/solarsoft/latest_events/}{Latest Events}" archive in 
SolarSoft\footnote{SolarSoft Latest Events: \url{http://www.lmsal.com/solarsoft/latest_events/}}
and are not discussed further.
Table~\ref{t-results} lists the event counts for each 
category, while Figures~\ref{f-type}-\ref{f-cme} provide representative examples 
and bar plots of the category distributions. All images are single frames from the event movies 
provided in the catalog with no additional processing. Clicking the ID numbers 
listed in the electronic text 
(e.g. \href{http://aia.cfa.harvard.edu/filament/index.html?search=0962}{No. 0962}) 
will launch the catalog website, where the associated movies 
can be found. 

\begin{table}
\caption{Properties and event counts. Ambiguous (e.g. ``Y?") event counts are listed in parens. }
\label{t-results}
\begin{tabular}{lc|c|cccccc}
\hline
& & \multicolumn{7}{c}{Type} \\
Category & Bin & All & AR & IP & QS & PC & TE & ?  \\
\hline
\multicolumn{2}{c|}{Type} & \TypeAll{} & \TypeARg{} (\TypeARq{}) & \TypeIPg{} (\TypeIPq{}) & \TypeQSg{} (\TypeQSq{}) & \TypePCg{} (\TypePCq{}) & \TypeTEg{} (\TypeTEq{}) & \TypeQg{}  \\
\hline
\multirow{3}{*}{Position} & Disk & \AllPosDg{}  & \ARPosDg{}  & \IPPosDg{}  & \QSPosDg{}  & \PCPosDg{}  & \TEPosDg{}  & \QPosDg{}  \\
 & Limb & \AllPosLg{}  & \ARPosLg{}  & \IPPosLg{}  & \QSPosLg{}  & \PCPosLg{}  & \TEPosLg{}  & \QPosLg{}  \\
 & Behind & \AllPosBg{}  & \ARPosBg{}  & \IPPosBg{}  & \QSPosBg{}  & \PCPosBg{}  & \TEPosBg{}  & \QPosBg{}  \\
\hline
\multirow{4}{*}{Eruption} & Full & \AllEruFg{} (\AllEruFq{}) & \AREruFg{} (\AREruFq{}) & \IPEruFg{} (\IPEruFq{}) & \QSEruFg{} (\QSEruFq{}) & \PCEruFg{} (\PCEruFq{}) & \TEEruFg{}  & \QEruFg{} (\QEruFq{}) \\
 & Partial & \AllEruPg{} (\AllEruPq{}) & \AREruPg{} (\AREruPq{}) & \IPEruPg{} (\IPEruPq{}) & \QSEruPg{} (\QSEruPq{}) & \PCEruPg{} (\PCEruPq{}) & \TEEruPg{}  & \QEruPg{} (\QEruPq{}) \\
 & Confined & \AllEruCg{} (\AllEruCq{}) & \AREruCg{} (\AREruCq{}) & \IPEruCg{} (\IPEruCq{}) & \QSEruCg{} (\QSEruCq{}) & \PCEruCg{} (\PCEruCq{}) & \TEEruCg{}  & \QEruCg{} (\QEruCq{}) \\
 & Other\tabnote{Indicates non-eruptive events that are excluded from all other counts.} & \AllEruOg{} (\AllEruOq{}) & \AREruOg{} (\AREruOq{}) & \IPEruOg{}  & \QSEruOg{} (\QSEruOq{}) & \PCEruOg{}  & \TEEruOg{}  & \QEruOg{}  \\
\hline
\multirow{3}{*}{Symmetry} & Sym & \AllSymSg{} (\AllSymSq{}) & \ARSymSg{} (\ARSymSq{}) & \IPSymSg{} (\IPSymSq{}) & \QSSymSg{} (\QSSymSq{}) & \PCSymSg{} (\PCSymSq{}) & \TESymSg{} (\TESymSq{}) & \QSymSg{} (\QSymSq{}) \\
 & Asym & \AllSymAg{} (\AllSymAq{}) & \ARSymAg{} (\ARSymAq{}) & \IPSymAg{} (\IPSymAq{}) & \QSSymAg{} (\QSSymAq{}) & \PCSymAg{} (\PCSymAq{}) & \TESymAg{} (\TESymAq{}) & \QSymAg{} (\QSymAq{}) \\
 & ? & \AllSymQg{}  & \ARSymQg{}  & \IPSymQg{}  & \QSSymQg{}  & \PCSymQg{}  & \TESymQg{}  & \QSymQg{}  \\
\hline
\multirow{4}{*}{Direction} & Radial & \AllDirRg{} (\AllDirRq{}) & \ARDirRg{} (\ARDirRq{}) & \IPDirRg{} (\IPDirRq{}) & \QSDirRg{} (\QSDirRq{}) & \PCDirRg{} (\PCDirRq{}) & \TEDirRg{} (\TEDirRq{}) & \QDirRg{} (\QDirRq{}) \\
 & Non-R & \AllDirNRg{} (\AllDirNRq{}) & \ARDirNRg{} (\ARDirNRq{}) & \IPDirNRg{} (\IPDirNRq{}) & \QSDirNRg{} (\QSDirNRq{}) & \PCDirNRg{} (\PCDirNRq{}) & \TEDirNRg{}  & \QDirNRg{} (\QDirNRq{}) \\
 & Sideways & \AllDirSg{} (\AllDirSq{}) & \ARDirSg{} (\ARDirSq{}) & \IPDirSg{} (\IPDirSq{}) & \QSDirSg{} (\QSDirSq{}) & \PCDirSg{} (\PCDirSq{}) & \TEDirSg{}  & \QDirSg{} (\QDirSq{}) \\
 & ? & \AllDirQg{}  & \ARDirQg{}  & \IPDirQg{}  & \QSDirQg{}  & \PCDirQg{}  & \TEDirQg{}  & \QDirQg{}  \\
\hline
\multirow{3}{*}{Twist} & Yes & \AllTwiYg{} (\AllTwiYq{}) & \ARTwiYg{} (\ARTwiYq{}) & \IPTwiYg{} (\IPTwiYq{}) & \QSTwiYg{} (\QSTwiYq{}) & \PCTwiYg{} (\PCTwiYq{}) & \TETwiYg{}  & \QTwiYg{} (\QTwiYq{}) \\
 & No & \AllTwiNg{} (\AllTwiNq{}) & \ARTwiNg{} (\ARTwiNq{}) & \IPTwiNg{} (\IPTwiNq{}) & \QSTwiNg{} (\QSTwiNq{}) & \PCTwiNg{} (\PCTwiNq{}) & \TETwiNg{} (\TETwiNq{}) & \QTwiNg{} (\QTwiNq{}) \\
 & ? & \AllTwiQg{}  & \ARTwiQg{}  & \IPTwiQg{}  & \QSTwiQg{}  & \PCTwiQg{}  & \TETwiQg{}  & \QTwiQg{}  \\
\hline
\multirow{3}{*}{Writhe} & Yes & \AllKinYg{} (\AllKinYq{}) & \ARKinYg{} (\ARKinYq{}) & \IPKinYg{} (\IPKinYq{}) & \QSKinYg{} (\QSKinYq{}) & \PCKinYg{} (\PCKinYq{}) & \TEKinYg{}  & \QKinYg{} (\QKinYq{}) \\
 & No & \AllKinNg{} (\AllKinNq{}) & \ARKinNg{} (\ARKinNq{}) & \IPKinNg{} (\IPKinNq{}) & \QSKinNg{} (\QSKinNq{}) & \PCKinNg{} (\PCKinNq{}) & \TEKinNg{}  & \QKinNg{} (\QKinNq{}) \\
 & ? & \AllKinQg{}  & \ARKinQg{}  & \IPKinQg{}  & \QSKinQg{}  & \PCKinQg{}  & \TEKinQg{}  & \QKinQg{}  \\
\hline
Vertical & Yes & \AllVerYg{} (\AllVerYq{}) & \ARVerYg{} (\ARVerYq{}) & \IPVerYg{} (\IPVerYq{}) & \QSVerYg{} (\QSVerYq{}) & \PCVerYg{} (\PCVerYq{}) & \TEVerYg{}  & \QVerYg{}  \\
Threads\tabnote{Vertical threads counts exclude events behind the limb.} & No & \AllVerNg{} (\AllVerNq{}) & \ARVerNg{} (\ARVerNq{}) & \IPVerNg{} (\IPVerNq{}) & \QSVerNg{} (\QSVerNq{}) & \PCVerNg{}  & \TEVerNg{} (\TEVerNq{}) & \QVerNg{}  \\
\hline
\multirow{3}{*}{Cavity\tabnote{Cavity counts include limb events only.}} & Yes & \AllCavYg{} (\AllCavYq{}) & \ARCavYg{} (\ARCavYq{}) & \IPCavYg{}  & \QSCavYg{} (\QSCavYq{}) & \PCCavYg{} (\PCCavYq{}) & \TECavYg{}  & \QCavYg{} (\QCavYq{}) \\
 & No & \AllCavNg{} (\AllCavNq{}) & \ARCavNg{}  & \IPCavNg{}  & \QSCavNg{} (\QSCavNq{}) & \PCCavNg{} (\PCCavNq{}) & \TECavNg{}  & \QCavNg{}  \\
 & During & \AllCavDg{} (\AllCavDq{}) & \ARCavDg{} (\ARCavDq{}) & \IPCavDg{} (\IPCavDq{}) & \QSCavDg{} (\QSCavDq{}) & \PCCavDg{}  & \TECavDg{}  & \QCavDg{} (\QCavDq{}) \\
\hline
\multirow{2}{*}{CME} & Yes & \AllCMEYg{} (\AllCMEYq{}) & \ARCMEYg{} (\ARCMEYq{}) & \IPCMEYg{} (\IPCMEYq{}) & \QSCMEYg{} (\QSCMEYq{}) & \PCCMEYg{} (\PCCMEYq{}) & \TECMEYg{}  & \QCMEYg{} (\QCMEYq{}) \\
 & No & \AllCMENg{} (\AllCMENq{}) & \ARCMENg{} (\ARCMENq{}) & \IPCMENg{} (\IPCMENq{}) & \QSCMENg{} (\QSCMENq{}) & \PCCMENg{} (\PCCMENq{}) & \TECMENg{}  & \QCMENg{} (\QCMENq{}) \\
\hline
\multirow{2}{*}{Flare} & Yes & \AllFlaYg{} (\AllFlaYq{}) & \ARFlaYg{} (\ARFlaYq{}) & \IPFlaYg{} (\IPFlaYq{}) & \QSFlaYg{} (\QSFlaYq{}) & \PCFlaYg{} (\PCFlaYq{}) & \TEFlaYg{}  & \QFlaYg{}  \\
 & No & \AllFlaNg{} (\AllFlaNq{}) & \ARFlaNg{} (\ARFlaNq{}) & \IPFlaNg{} (\IPFlaNq{}) & \QSFlaNg{} (\QSFlaNq{}) & \PCFlaNg{} (\PCFlaNq{}) & \TEFlaNg{}  & \QFlaNg{}  \\
\hline
\end{tabular}
\end{table}

\subsection{Type}
\label{type}

We assign types based on the commonly-used, location-based scheme. \textit{Active region}
(AR) filaments are found within active regions, \textit{intermediate} prominences (IP) are 
found adjacent to or between active regions, \textit{quiescent} (QS) filaments are found well 
away from active regions in the quiet sun, and \textit{polar crown} (PC) prominences are 
found at high latitudes. 
Figure~\ref{f-type} shows representative examples of each type.
Note that PC may refer specifically to filaments 
that form at the boundaries of the polar coronal holes, but we use the term more 
generally to refer to any filament with a latitude greater than $\sim$50$^\circ$ N/S.
Including ambiguous events (e.g. ``AR?"), our sample includes 
 \TypeAR{} AR, \TypeIP{} IP, \TypeQS{} QS, and \TypePC{} PC filament eruptions. 
 An additional \TypeQ{} eruptions are of indeterminable type.
 These events are generally behind the limb without available STEREO observations, 
 though a few are very compact filaments in small emerging flux regions 
 that do not fit neatly into the four main groups. 
 \TypeTE{} more are classified as \textit{transequatorial} (TE) filaments, the longest 
 and rarest variety. Neither of these latter two groups are included in the discussions 
 below, but we find that the TE events have characteristics most similar to QS eruptions. 
Other classification schemes exist based on morphology, 
dynamics, spectroscopic characteristics, and magnetic configuration. A review of these, 
along with historical classifications, is provided by \citet{Engvold15}. It would 
be worthwhile to investigate a subset of the events and characteristics described below using 
additional type sets, particularly the scheme based on photospheric flux distribution 
advanced by \citet{Mackay08}. 

The extent to which each filament erupts is also categorized, guided by 
conventions described by \citet{Gilbert07}. 
A \textit{full} (F) eruption is one in which the bulk of the filament mass 
escapes the AIA field-of-view (FOV) along with the full magnetic structure,  
a \textit{partial} (P) eruption is one in which the magnetic structure erupts either 
completely or partially, expelling some or none of the filament mass, 
and a \textit{confined} (C) or failed eruption is one in which none of the 
filament mass nor magnetic structure escapes. 
It can be difficult in practice to apply these straightforward definitions 
because all events exhibit some draining, whereby expelled    
material flows back down the legs of an erupting prominence. 
This returned mass may constitute an apparently significant volumetric fraction 
of the material seen, but it is not necessarily clear what fraction 
of the actual mass escapes given that the volume of the structure 
is evolving and that the 304 \AA{} observations 
we primarily rely on are optically thick even at low column densities. 
For this reason, we relax the $>$90\% criterion used by \citet{Gilbert07} 
and define a full eruption as one in which the magnetic structure erupts 
completely and no more than a third of the filament material 
appears to drain back down. Question marks are used to denote 
events near this boundary. 

It can also be difficult to make a determination  
for faint events that become too diffuse to be tracked to the edge of the 
FOV or distinguished from the background chromosphere on the disk; these 
are labeled as full if they appeared to be erupting fully 
prior to becoming untraceable. 
The very few instances in which 
a filament channel erupts with little or no filament material 
contained within it are labeled as partial eruptions 
(e.g. \href{http://aia.cfa.harvard.edu/filament/index.html?search=0478}{No. 0478}). 
Finally, note that these labels are of course based on the behavior of the 
filament and associated magnetic structure to the extent 
to which they are visible in the three passbands we examine. 
It may be possible, for instance, for an apparently confined event 
to produce a detectable CME either because an eruption was 
stimulated in the overlying field or because some part of the 
prominence structure escaped without producing a noticeable signature 
in our observations (see \S\ref{cmes}).
Given these definitions and including ambiguous events (e.g. F?), the 
catalog includes \AllEruF{} full, \AllEruP{} partial, and \AllEruC{} confined eruptions. 
An additional \AllEruO{} events are labeled as \textit{other} (O), which include non-eruptive dynamics 
such as filament activation. These events are included in the online catalog but are 
excluded from all counts in this paper. 


\subsection{Symmetry}
\label{symmetry}

 \begin{figure}
 \centerline{\includegraphics[width=\textwidth,clip=]{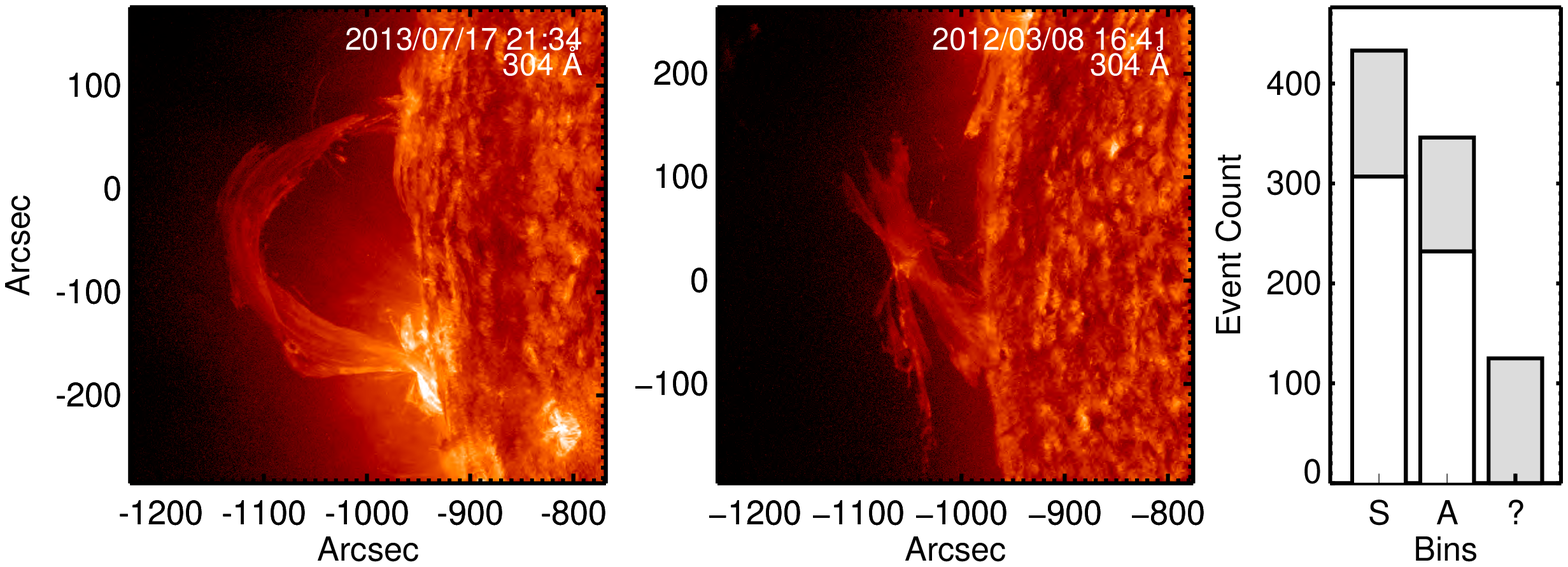}}
 \caption{Examples of symmetric 
 (left, \href{http://aia.cfa.harvard.edu/filament/index.html?search=0745}{No. 0745}) 
 and asymmetric 
 (middle, \href{http://aia.cfa.harvard.edu/filament/index.html?search=0405}{No. 0405}) eruptions. The right panel shows the number of 
 events found for each type, and shaded regions denote ambiguous events (e.g. ``S?").}
 \label{f-symmetry}
 \end{figure}

Filament eruptions are often divided into symmetric and asymmetric categories depending on whether 
they exhibit a loop-like geometry or one that favors the liftoff of a single footpoint. This distinction is particularly useful 
for studies of potential eruption mechanisms and magnetic configurations.
For instance, \citet{Tripathi06} examined 
EUV brightenings associated with both types and found that in the symmetric case, brightenings propagated from the 
middle to each end, while in the asymmetric case, they moved from the erupting end toward the tethered one. 
Based on this, they suggest 
a three-dimensional extension to the standard two-dimensional flare model in which the EUV brightenings are caused 
by successive reconnection events. \citet{Liu09} further divided asymmetric events into 
``whipping" and ``zipping" types for which the active leg either whips upward dramatically or moves along the polarity 
inversion line like the unfastening of a zipper, with each case exhibiting distinct hard X-ray patterns and 
implying different field configurations.  

We distinguish simply between the \textit{symmetric} (S) and \textit{asymmetric} (A) cases based primarily on inspection of 304 \AA{} observations. 
This distinction may be obscured  
by the orientation of the filament and its two-dimensional projection, such as when the filament has a predominantly 
east-west axis and is positioned on the limb. For this reason, there are a large number of ambiguous 
(e.g. ``S?") and indeterminable (``?") events. 
Figure~\ref{f-symmetry} provides examples of both cases along with a bar plot showing the frequency distribution. 
Including ambiguous events, 
we find that \AllSymSp{} of eruptions are symmetric, \AllSymAp{} are 
asymmetric, and \AllSymQp{} are indeterminable. The respective percentages of symmetric and asymmetric events 
by filament type are \ARSymSp{} and \ARSymAp{} for AR, \IPSymSp{} and \IPSymAp{} for IP, 
\QSSymSp{} and \QSSymAp{} for QS, and \PCSymSp{} and \PCSymAp{} for PC eruptions. Symmetric 
eruptions are thus somewhat more common for each type except polar crown filaments. 
It is possible that the longer length of PC filaments could provide more opportunities for one 
footpoint to be preferentially destabilized by neighboring events, however these differences 
may not be very significant given large number of questionable and indeterminable events. 
More significantly, symmetric events are around 1.5$\times$ more likely 
to be full eruptions than asymmetric ones. This is primarily because 
the asymmetric geometry allows mass to more easily drain out 
of the filament before it is fully expelled and because asymmetric events 
frequently leave behind a section of the filament at the 
end opposite to the initial liftoff. 


\subsection{Direction}
\label{direction}

 \begin{figure}
 \centerline{\includegraphics[width=\textwidth,clip=]{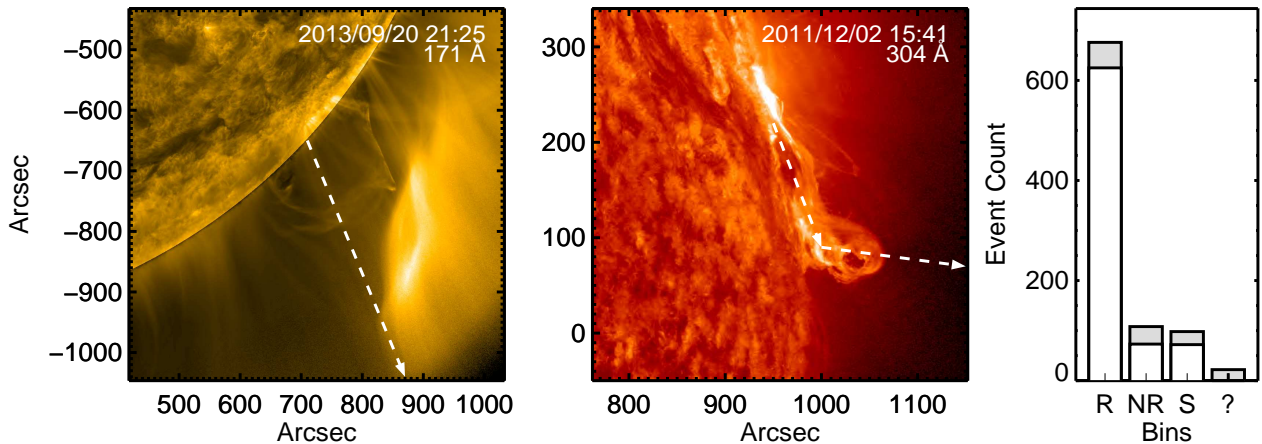}}
 \caption{Examples of non-radial 
 (left, \href{http://aia.cfa.harvard.edu/filament/index.html?search=0797}{No. 0797}) 
 and sideways 
 (middle, \href{http://aia.cfa.harvard.edu/filament/index.html?search=0332}{No. 0332}) 
 trajectories. Arrows denote the approximate trajectories. The right 
 panel shows the total number of radial (R), sideways (S), and non-radial (NR) events, 
 and the shaded regions denote ambiguous events (e.g. ``NR?").}
 \label{f-direction}
 \end{figure}
 
 The directions of filament eruptions and associated CMEs are influenced by the surrounding magnetic environment, which 
 may deflect an eruption away from a simple radial trajectory. Such interactions between small- and large-scale 
 magnetic structures are of particular interest in the space weather context because CME deflection influences an event's potential  
 to impact Earth \citep{Kay14}. \citet{Gopalswamy15} provides a recent review of non-radial prominence trajectories, 
 which generally match that of the associated white-light CMEs \citep{Simnett00} but often exhibit a greater degree 
 of non-radiality \citep{Panasenco13}. Eruptions will tend to travel along the paths of least resistance, deflecting away from regions 
 of higher magnetic energy density \citep{Gui11} such as coronal holes and active regions. 
 
We distinguish between three types of trajectories: \textit{radial} (R), \textit{non-radial} (NR), and \textit{sideways} (S). 
Non-radial events are somewhat inclined from a radial trajectory and are generally deflected 
by large scale features such as coronal holes and pseudo-streamers. Sideways events have a 
substantial tangential component initially and are generally deflected away from active regions, often  
adopting a radial trajectory while still within the AIA FOV. 
Figure~\ref{f-direction} provides examples of the latter two cases along with a bar plot showing 
the category distributions. 
The non-radial example is deflected southward by an overlying pseudo-streamer and the sideways 
example is ejected tangentially out of its host AR. 
Including ambiguous 
events (e.g. ``NR?"), we find that \AllDirRp{} of eruptions are radial, \AllDirNRp{} are non-radial, 
\AllDirSp{} are sideways, and \AllDirQp{} are indeterminable. By type, the respective percentages 
of non-radial and sideways events are \ARDirNRp{} and \ARDirSp{} for AR, 
\IPDirNRp{} and \IPDirSp{} for IP, \QSDirNRp{} and \QSDirSp{} for QS, 
and \PCDirNRp{} and \PCDirSp{} for PC eruptions. 
Similar non-radial rates of around 12 $\pm$ 1\% are found for full, partial, and confined eruptions. 
However, these groups exhibit sideways trajectories at different rates of 7\%, 
12\%, and 16\%, respectively. 
Sideways eruptions are more frequently confined or partial perhaps because they are 
more likely to encounter neighboring magnetic structures that 
might arrest the entire eruption or divert some ejecta along a 
different set of field lines. Alternatively, eruptions may be confined 
due to the prevailing strength of the overlying field and recorded 
as sideways if the material is diverted tangentially before draining back to the surface.

Future work on quantifying the extent to which eruptions 
deviate from the radial direction and on what prompts this deviation might start from the events identified here. 
In particular, there is at present no clearly-articulated model describing the evolution of the magnetic 
field in the sideways ejection events from ARs. One hypothesis is that such trajectories arise  
from configurations in which the overlying coronal arcade in the center of the 
active region is too strong for the flux rope to break through radially. Instead, 
the system may first destabilize in the weaker-field region at the active region periphery. 
This might explain why filament plasma is seen to move 
horizontally from strong to weak field regions at the start of some events. 
However, if the stronger-field side is suddenly diminished by flux cancellation or 
footpoint motions, the filament may also eject toward the weakened stronger-field side. 
Thus, new insights might be gleaned from magnetic field modeling along with a  
careful examination of potential changes in the photospheric 
flux distribution for some events in our sideways sample. 


\subsection{Twist}
\label{twist}

 \begin{figure}
 \centerline{\includegraphics[width=\textwidth,clip=]{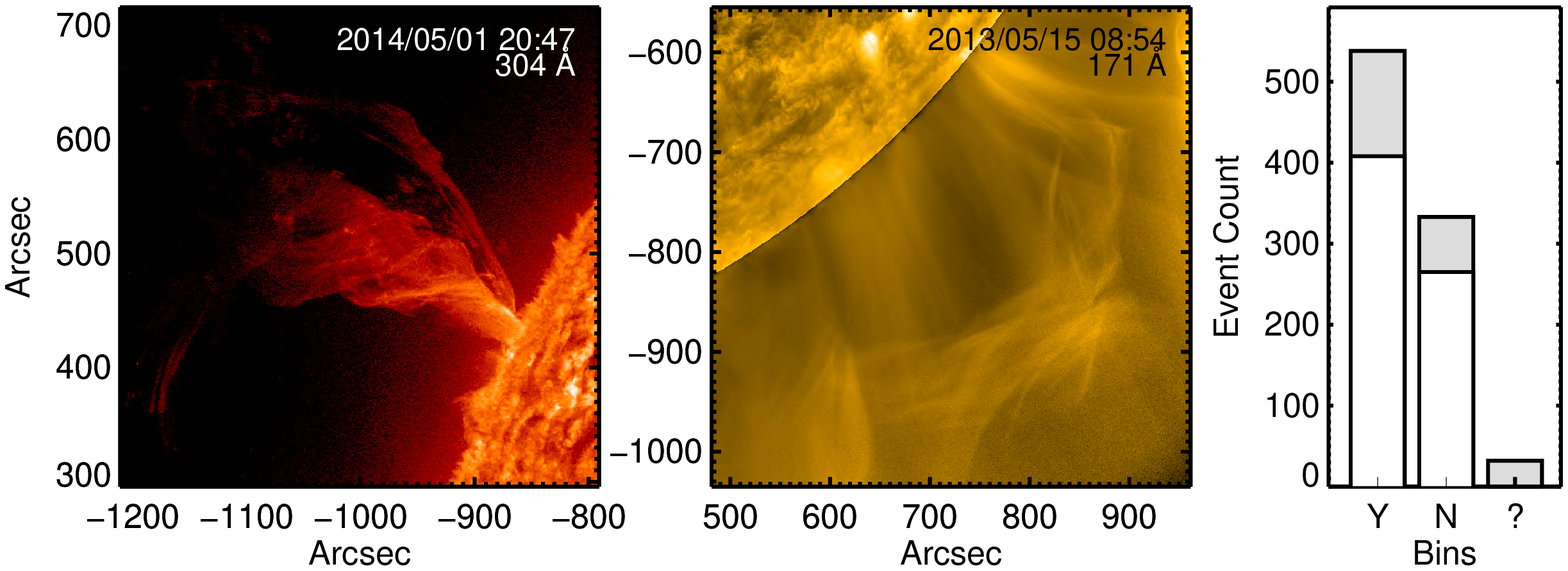}}
 \caption{Examples of events exhibiting twist
 (left: \href{http://aia.cfa.harvard.edu/filament/index.html?search=0994}{No. 0994}, 
 right: \href{http://aia.cfa.harvard.edu/filament/index.html?search=0696}{No. 0696}). 
 The right panel shows the number 
 of events that do (Y) and do not (N) show twist, and the shaded regions denote 
 ambiguous events (e.g. ``Y"?).}
 \label{f-twist}
 \end{figure}

Apparent twist in filament eruptions is particularly intriguing  
because of its relationship to stored energy release and interactions with nearby magnetic structures. 
Several types of twist may be observed, each with different implications. For instance, 
the ``roll effect" refers to rotation about the filament axis whereby 
the loop apex rolls to one side, creating twist of opposite signs in the two legs \citep{Martin03}. 
\citet{Panasenco11} explore this effect for three eruptions observed 
by STEREO, and \citet{Panasenco13} consider its connection to prominence 
trajectories, showing a relationship between roll and the presence of large-scale 
features like coronal holes and pseudo-streamers that may induce rolling motions through deflection. 
\citet{Murphy12} suggest 
that rolling may also arise from an offset between a rising 
flux rope and the corresponding CME current sheet. 

\citet{Su13} model a PC 
eruption, reproducing rolling motions observed at low heights 
through a continuous injection of twist from the active region side that 
creates an asymmetric field distribution, 
possibly inducing rolling motions through asymmetric reconnection
similar to that described by \citet{Murphy12}. 
\citeauthor{Su13} also describe further rotating motions attributed to 
the relaxing of dipped field lines after the eruption (see also \citealt{Thompson13}). 
\citet{Romano03} explore  
untwisting motions in detail, which can often be seen in one or both legs 
during eruptions (e.g. \citealt{Yan14}). 
Mass flow during an eruption may also reveal 
twisted structure inherent to the filament. 
Finally, a distinction can be drawn between 
\textit{twist}, which generally refers to the degree to which field lines are wound around a 
central axis, and \textit{writhe}, the helical deformation (twist) of the axis itself  
that may lead to the kink instability \citep{Torok10}. These properties  
may be combined into the mathematical measure \textit{helicity}, which 
quantifies the degree to which a magnetic field is twisted, writhed, and linked. 
Helicity is positively related to free magnetic energy, 
and helicity accumulation is thought to be very important in leading to an eruption \citep{Tziotziou14}.

In this work, we assess simply whether (un)twisting motions of any type are apparent in each eruption, 
and references to ``twisting motions" here and in later sections indicate the presence of at least one 
of the twist signatures described above. 
It was our initial intent to subdivide twist into categories related to the previous discussion, 
but we found that the careful inspection required to make these distinctions was untenable 
for our large sample and additional aims. Future work on this topic is planned, and some 
preliminary results were shown by \citet{McKillop14}.
However, we do note that the untwisting motions in filament legs 
and mass flow that reveals twisted structure are the 
most commonly-observed of these phenomena in our events. 
Figure~\ref{f-twist} shows two examples and the event count distribution. 
Including ambiguous events (e.g. ``Y?"), we find that \AllTwiYp{} of 
events show signs of twist, \AllTwiNp{} do not, and the remainder 
are indeterminable. By type, the percentage of events with evident twist 
are \ARTwiYp{} for AR, \IPTwiYp{} for IP, 
\QSTwiYp{} for QS, and \PCTwiYp{} for PC eruptions. We therefore find that a majority of 
all filament types exhibit signs of twist and that the likelihood is highest for 
IP eruptions. 

We also find that events with apparent twisting motions have 
faster CME speeds on average, 
and that IP events are generally the fastest of the four types (see 
\S\ref{results} and \S\ref{speed_twist}).
Full and partial eruptions exhibit twisting motions at similar 
rates of 61\% and 66\%, respectively. 
Interestingly, confined eruptions are significantly less likely to 
exhibit signs of twist, with a rate of 44\%, 
despite including a comparatively higher rate of 
events with writhing motions (see \S\ref{kink}). 
We hypothesize that events exhibiting twisting motions will have had 
greater helicity on average, which implies 
a greater release of magnetic energy during the eruptions. 
The lower rate of twisting motions in confined events thus might  
be understood in terms of the energy available to push through 
the overlying field.  
Relationships between apparent twisting motions and eruption kinematics 
will be explored in \S\ref{speed_twist}.


\subsection{Writhe}
\label{kink}

 \begin{figure}    
 \centerline{\includegraphics[width=\textwidth,clip=]{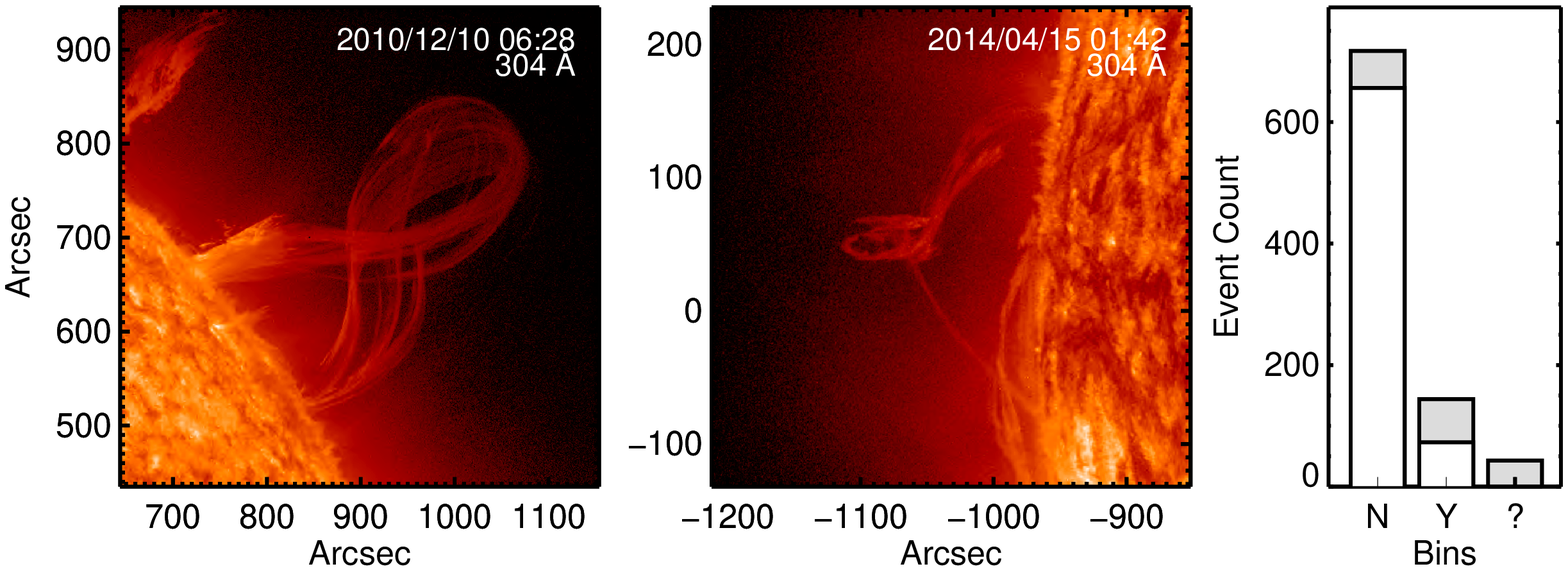}}
 \caption{Examples of events exhibiting writhe 
 (left: \href{http://aia.cfa.harvard.edu/filament/index.html?search=0126}{No. 0126}, 
 right: \href{http://aia.cfa.harvard.edu/filament/index.html?search=0927}{No. 0927}). 
 The right panel shows the number 
 of events that do (Y) and do not (N) show writhing, and the shaded regions denote 
 ambiguous events (e.g. ``Y"?).}
 \label{f-kink}
 \end{figure}
 
 The helical kink instability is an ideal magnetohydrodynamic (MHD) instability that affects columns of plasma 
 with strong axial currents, such as magnetic flux ropes, and is so named for the characteristic
 inverted gamma shape of kinked rope. 
 As described in the previous section, the helical deformation of a flux rope's axis that may ultimately 
 produce a kinked shape is generally referred to as \textit{writhe} \citep{Torok10}.
 The kink instability is one of the principal 
 mechanisms thought to drive prominence eruptions (see \S\ref{discussion} 
 and review by \citealt{Fan15}), 
 and numerical simulations of kink-unstable flux ropes have been very successful 
 in reproducing characteristics of observed events \citep{Torok05}, suggesting
that coronal flux ropes exist prior to eruption \citep{Gibson06b}. An overview 
of how the kink instability manifests in solar observations is given by \citet{Gilbert07}, and a
recent effort to quantify the twist in a likely kink-unstable event is presented by \citet{Yan14}.

We note the appearance of writhing based on inspection of the 304 and 171 \AA{} observations. 
Figure~\ref{f-kink} provides two examples along with a bar plot showing the distribution of 
events with and without.
Including ambiguous events (e.g. ``Y?"), we find that \AllKinNp{} of eruptions show no writhe, 
\AllKinYp{} do, and another \AllKinQp{} were indeterminable. However, fully 
half of the \AllKinY{} writhed events are labeled as ambiguous. This is due to the difficulty in positively  
identifying the inverse-gamma structure, which may be obscured by 
the orientation or dynamics of the event. By type, we find that \ARKinYp{} of AR, \IPKinYp{} of IP, 
\QSKinYp{} of QS, and \PCKinYp{} of PC eruptions exhibit writhing motions. 
62\% and 23\% of the writhed events are partial and confined eruptions, respectively, which are 
significantly greater than the general population rates of 40\% and 20\%. 
This is consistent with expectations from \citet{Gilbert07}, who note that 
kinking structures are highly unlikely to produce full eruptions 
because the filament material needs to be concentrated 
at the center of the flux rope and the eruption needs to be sufficiently 
fast to limit draining. We do find that 15\% of our kinked sample (22 events) 
are full eruptions, which might be greater than expected by \citeauthor{Gilbert07}.  
This is largely due to our comparatively lax definition of ``full" (see \S\ref{type}), 
which permits more draining,
however at least a few events exhibiting writhing motions appear to be full eruptions 
even with the stricter definition 
(e.g. \href{http://aia.cfa.harvard.edu/filament/index.html?search=0770}{No. 0770} \& 
\href{http://aia.cfa.harvard.edu/filament/index.html?search=0207}{No. 0207}).
It is also important to note that a number of additional events classified as having 
apparent twisting motions in the previous section might also have been 
influenced by the kink instability but had underdeveloped or obscured 
inverse-gamma structures.


\subsection{Vertical Threads}
\label{threads}

 \begin{figure}    
 \centerline{\includegraphics[width=\textwidth,clip=]{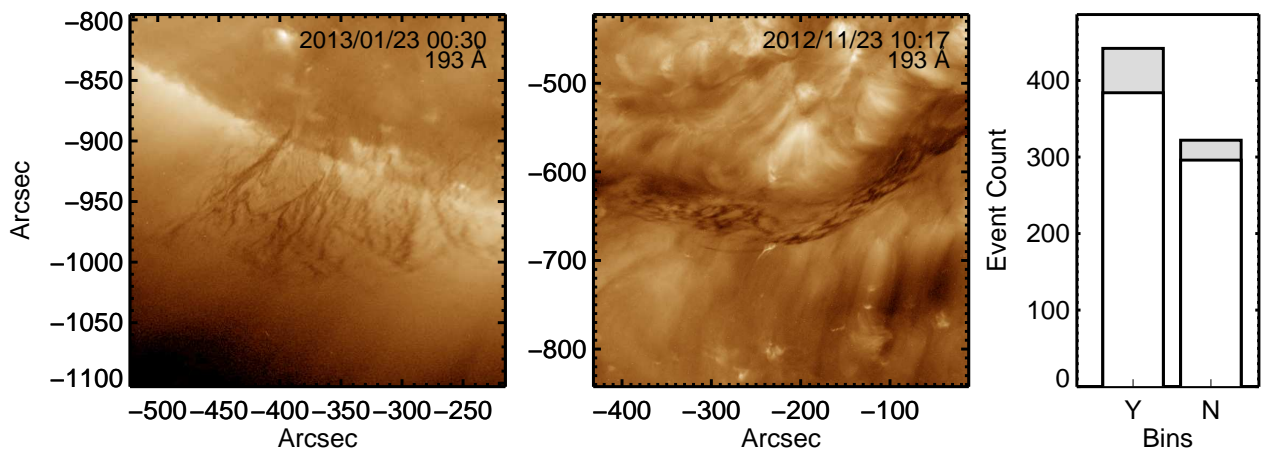}}
 \caption{Examples of vertical absorption threads on the limb 
 (left, \href{http://aia.cfa.harvard.edu/filament/index.html?search=0578}{No. 0578}) 
 and disk 
 (middle, \href{http://aia.cfa.harvard.edu/filament/index.html?search=0797}{No. 0797}). 
  The right panel shows the number of events with (Y) and without (N) vertical threads, excluding 
  events behind the limb,  
  and the shaded regions denote ambiguous events (e.g. ``Y?").
  }
 \label{f-threads}
 \end{figure}
 
 Thin threads, revealed primarily by high-resolution H$\alpha{}$ imaging, are the ubiquitous building 
 blocks of solar filaments  \citep{Lin08}. These are field-aligned and oriented horizontally, parallel 
 or somewhat inclined with respect to filament spines and barbs. 
 In addition to horizontal threads, 
 vertical threads, perpendicular to the spine, are frequently observed in quiescent and polar 
 crown ``hedgerow" prominences \citep{Engvold15}. Vertical threads are 
 challenging to explain given the predominantly horizontal orientation of the local magnetic 
 field \citep{Paletou01}, and several scenarios have been proposed. For instance, \citet{Ballegooijen10} 
 suggest that vertical threads form in a tangled field environment, \citet{Chae10} suggests that they form from stacks 
 of sagging horizontal threads, and \citet{Low12a,Low12b} suggest that vertical structures may arise from cross-field  
 mass transport after a breakdown of the frozen-in magnetic field condition, perhaps in concert with 
 magneto-thermal convection as described by \citet{Berger11}.
 
We note the visibility of vertical threads (or more generally, threads perpendicular to the filament spine) based 
primarily on inspection of the 193 \AA{} images. The threads appear dark in the EUV, as emission is 
absorbed and not produced by the neutral hydrogen and helium of the partially-ionized prominence plasma, which 
typically has $\rm{HI}$ column densities of $\sim$10\tsp{19} cm\tsp{-2} \citep{Labrosse11}. 
Figure~\ref{f-threads} shows examples of vertical threads seen on the limb and on the disk, along with a 
bar plot showing the number of events with and without. 
Including the ambiguous events (e.g. ``Y?"), and excluding those behind the limb, 
we find that \AllVerYp{} of events exhibit 
vertical threads and \AllVerNp{} do not. By type, the percentage 
of events with vertical threads are \ARVerYp{} for AR, 
\IPVerYp{} for IP, \QSVerYp{} for QS, and \PCVerYp{} for PC eruptions. 
The size of a filament influences whether the threads themselves or their orientations will 
be distinguishable, and we find rates that grow with the characteristic size scales of the four filament types. 
Confined events are also much less likely to exhibit vertical threads (43\%), perhaps
due to their generally smaller sizes.  
This may be caused by limited spatial resolution, or it may be that compact filaments 
are indeed less likely to contain vertical threads.
We also note that the \ARVerY{} AR events that do exhibit vertical threads 
are generally more extended, with higher pre-eruption heights, 
than average for that type. Several of these events are also on the boundary between 
what might be considered an ``active region" versus ``intermediate" filament or are 
found within decaying active regions with well-separated polarities.     
Thus, we find the presence of vertical threads to be the norm outside of active regions, and we note 
that our rates may be underestimated because some small or unfavorably-oriented 
filaments were likely excluded. 


\subsection{Cavity}
\label{cavity}

  \begin{figure}    
 \centerline{\includegraphics[width=\textwidth,clip=]{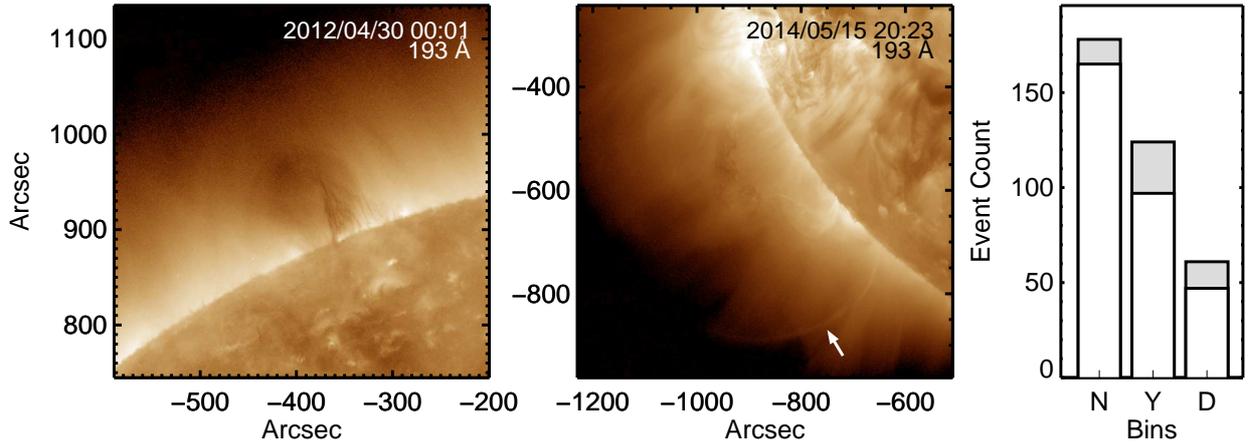}}
 \caption{Examples of a coronal cavity 
 (left, \href{http://aia.cfa.harvard.edu/filament/index.html?search=0452}{No. 0452}) 
 and a cavity that became apparent during the eruption 
 (middle, see movie: \href{http://aia.cfa.harvard.edu/filament/index.html?search=0953}{No. 0953}).
 The right panel shows the number of \textit{limb} events with (Y), without (N), 
 and during (D), and the shaded regions denote ambiguous events (e.g. ``Y?"). 
 }
 \label{f-cavity}
 \end{figure} 
 
 Coronal cavities are elliptical depressions in the intensities above and surrounding 
 prominences observed at optical, EUV, and soft X-ray wavelengths. They can be  
 understood as the limb projections of filament channels on the disk, which 
 may or may not contain a filament, and their interiors are less dense \citep{Fuller09} 
 and often hotter \citep{Reeves12} than the surrounding plasma. Cavities are 
 commonly modeled as magnetic flux ropes (e.g. \citealt{Low95,Fan12}) and may be inherent 
 components of filament systems that form from helicity accumulation \citep{Mackay15}.
 At least a third of cavities 
 erupt \citep{Forland13}, though many of those that do not may simply erupt sometime 
 after they were observed at the limb. CMEs often exhibit 
 a three-part structure that includes a bright plasma pileup followed by a dark cavity 
 and bright prominence core \citep{Hundhausen99}. A recent review 
 of cavities is provided by \citet{Gibson15}.

We note the visibility of associated cavities based on inspection of 193 \AA{} 
observations, which are sensitive to material around 1.2 MK in non-flaring regions. 
Of the AIA passbands, cavities are most clearly delineated in this channel along with the 211 \AA{} 
band, which is not included in the catalog. \textit{No} (N) indicates that a cavity was not apparent, 
\textit{yes} (Y) indicates that a cavity was visible 
prior to eruption, and \textit{during} (D) indicates that a cavity became visible during the eruption. 
Figure~\ref{f-cavity} 
provides two examples along with a bar plot of the results for all \AllPosL{} limb events in the catalog.
Including the ambiguous events (e.g. ``Y?"), we find that  
\AllCavNp{} have no obvious cavity prior to eruption, \AllCavYp{} do, and \AllCavDp{} have cavities that become visible 
after the eruption onset. 

By filament type, we find that cavities are apparent before and during the 
eruptions for \ARCavYp{} and \ARCavDp{} of AR, \IPCavYp{} and \IPCavDp{} of IP, 
\QSCavYp{} and \QSCavDp{} of QS, and \PCCavYp{} and \PCCavDp{} of PC eruptions, respectively. 
It is important to note that cavities are most apparent when the filament is large, oriented along 
the line-of-sight, and free of neighboring structures, so higher rates are expected 
for PC and QS prominences. Conversely, the compactness and proximity to active region arcades of AR 
and IP prominences explains why associated cavities are less frequently seen prior to eruptions and more frequently 
seen during eruptions, when motion makes the coherent structures more apparent. Finally, we note that 
additional processing such as radial filtering \citep{Forland13} can improve 
cavity visibility, which may have resulted in additional detections if applied to our 193 \AA{} observations. 
  

\subsection{Coronal Mass Ejections (CMEs)}
\label{cmes}

 \begin{figure}    
 \centerline{\includegraphics[width=\textwidth,clip=]{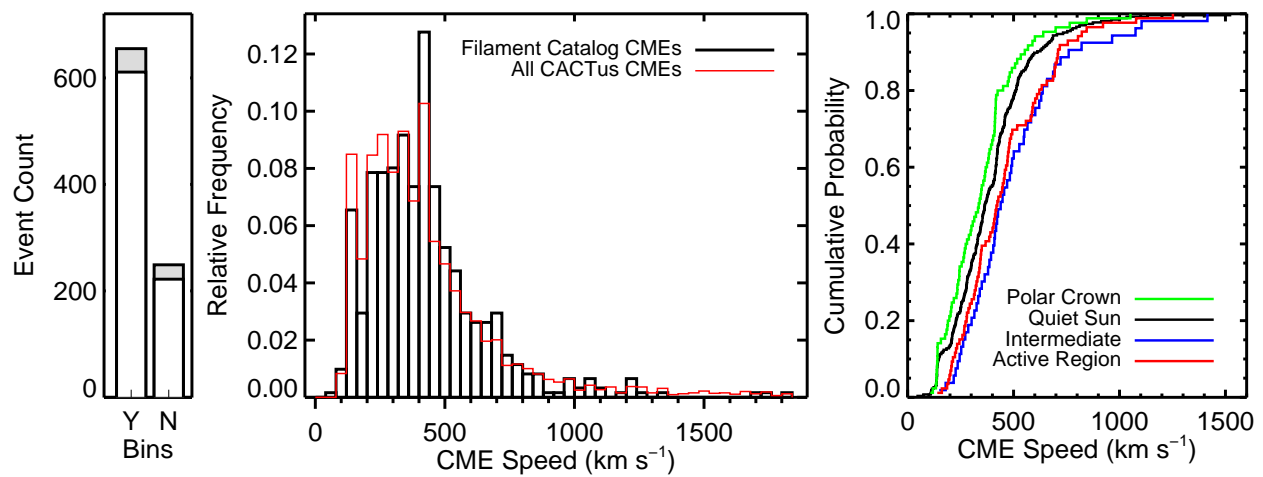}}
 \caption{
\textit{Left}: Distribution of events with and without associated white-light CMEs.
\textit{Middle}: Overall CME speed distribution.
\textit{Right}: Cumulative CME speed distributions by type.
 }
 \label{f-cme}
 \end{figure}
 
Coronal Mass Ejections (CMEs) are large expulsions of plasma and magnetic fields that 
escape into the heliosphere. A strong association between CMEs and filament eruptions has 
been known for many years from high statistical associations ($\sim$70\%; \citealt{Munro79}) and H$\alpha$ 
emission in CME cores \citep{Sheeley80}. As described in the previous section, CMEs often 
exhibit a three-part structure that includes the overlying prominence cavity and embedded filament. 
A review of CME and prominence eruption associations is provided by \citet{Webb15}. 
Complementary reviews on prominence eruption dynamics and space weather implications are 
also provided by \citet{Gopalswamy15} and \citet{Lugaz15}, respectively.
White-light coronagraphs have historically been the primary means by which CMEs are detected, and our 
associations are based on observations from the Large Angle Spectroscopic 
Coronagraph (LASCO; \citealt{Brueckner95}) aboard the Solar and Heliospheric 
Observatory (SOHO; \citealt{Domingo95}). Our definition of ``CME" is thus 
instrument-dependent, but it is important to note that the term refers to a 
physical process that is not limited to coronagraph observations. 

We search for matching events using the 
Computer Aided CME Tracking catalog
(CACTus\footnote{CACTus CME catalog: \url{http://sidc.oma.be/cactus/}}), 
which automatically detects CMEs, measuring both 
velocity and angular size. 
The speeds are averages of linear 
fits to radial height-time profiles that span the angular width of the CME and extend from 
1.5 to 30 \rsolar{} in height above the limb \citep{Robbrecht04}. 
These measurements reflect all parts of the CME, but note that 
there is variation within the three-part CME structure, with leading edges 
tending to be 30 to 40\% faster than core speeds \citep{Gopalswamy03, Mari09}.
In the few cases where LASCO or CACTus data are unavailable, 
we note the likelihood of a CME in the catalog with a question mark (``e.g. Y?").
We find that \AllCMEYp{} of our eruptions have associated CMEs 
and \AllCMENp{} do not. 
Similar rates are found when the events are divided by type, with CME-associations found for
\ARCMEYp{} of AR, \IPCMEYp{} of IP, 
\QSCMEYp{} of QS, and \PCCMEYp{} of PC eruptions. 

Basic statistics are listed in Table~\ref{t-cme} for several groupings, and  
Figure~\ref{f-cme} shows the CME speed distributions, which exhibit 
lognormal profiles that have been reported in other works \citep{Yurchyshyn05,Zhang06, Bein11}. 
We will see in \S\ref{kinematics} that this pattern emerges in the low corona. 
The average speed of CMEs associated with our filament eruptions is 430 \kms{}, which is  
essentially the same as for all CMEs detected by CACTus over the SDO mission lifetime. 
\citet{Moon02} compare similarly-observed CMEs associated with 
flares versus filament eruptions. They found a similar average velocity for 
filament-eruption CMEs but also reported these to be $\sim$9\% slower 
than the general population, which is not found here. This discrepancy may be due to 
the inclusion of false-positives in the CACTus catalog from transients 
related to previous CMEs or intensity variations in the slow solar wind that 
bias the total CACTus distribution toward lower speeds \citep{Robbrecht09}.

 
\begin{table}
\caption{CACTus CME Speeds}
\label{t-cme}
\begin{tabular}{l|c|cccc}
\hline
& & \multicolumn{4}{c}{CME Speed (\kms{})} \\
Category\tabnote{Subgroups exclude ambiguous events (e.g. Type = AR?).} & Events & Min & Max & Mean $\pm$ SD\tabnote{SD = Standard Deviation \& MAD = Median Absolute Deviation.} & Median $\pm$ MAD$^b$ \\
\hline
All CACTus & 6368 & 91 & 1950 & 428 $\pm$ 277 & 367 $\pm$ 182 \\
All Filament & 609 & 41 & 1840 & 425 $\pm$ 228 & 390 $\pm$ 158 \\
\hline
Active Region & 85 & 140 & 1320 & 466 $\pm$ 219 & 425 $\pm$ 163 \\
Intermediate & 52 & 149 & 1710 & 515 $\pm$ 272 & 449 $\pm$ 182 \\
Quiescent & 266 & 41 & 1840 & 391 $\pm$ 204 & 365 $\pm$ 143 \\
Polar Crown & 85 & 107 & 1220 & 351 $\pm$ 184 & 337 $\pm$ 133 \\
Full & 264 & 112 & 1740 & 461 $\pm$ 225 & 422 $\pm$ 157 \\
Partial & 210 & 41 & 1230 & 390 $\pm$ 208 & 361 $\pm$ 151 \\
Twist = Yes & 306 & 108 & 1740 & 432 $\pm$ 233 & 380 $\pm$ 160 \\
Twist = No & 156 & 41 & 1840 & 380 $\pm$ 210 & 374 $\pm$ 147 \\
\hline
\end{tabular}
\end{table}

We find, predictably, that full eruptions produce faster CMEs 
than partial eruptions by a factor of 1.2 on average. Full and partial 
eruptions have corresponding LASCO CMEs 95\% and 81\% of 
the time, respectively. Those that do not produce CMEs are 
generally very small, perhaps becoming confined or too 
diffuse to be detected at larger heights, or are oriented on the disk 
such that the opening angle is unfavorable for viewing by the coronagraph. 
Conversely, confined events lack CMEs 96\% of the time. The 
remaining 4\% either stimulate an eruption in a higher or neighboring structure,  
or the escaping filament material and/or magnetic 
structure was not perceptible in the AIA observations. 

We also find that intermediate prominence eruptions are the most likely 
to be associated with CMEs and that their CMEs tend to be the fastest. 
On average, IP-CMEs are 1.1$\times$ faster than AR-, 1.3$\times$ faster than 
QS-, and 1.5$\times$ faster than PC-CMEs. The cumulative probability 
distributions shown in the right panel of Figure~\ref{f-cme} indicate that 
the AR and IP curves are quite close. 
To test if their difference is statistically significant, we 
apply the k-sample Anderson-Darling ($A^2_k$) test,  
a non-parametric test that measures the deviation
between an arbitrary number 
of empirical distribution functions (2 in this case) and provides  
a significance level for which the null hypothesis that 
the groups are drawn from the same distribution can be rejected \citep{Scholz87}. 
The $A^2_k$ statistic for the IP and AR distributions is -0.50, corresponding to  
a $p$-value of 0.59, well above the 0.05 
level below which the distributions can be considered significantly 
different. We conclude, then, that our finding of enhanced 
CME speeds for IP versus AR eruptions is suggestive
but not statistically significant. 
We do find a significant relationship between apparent 
twisting motions and CME speed, which will be discussed in \S\ref{speed_twist}. 







\section{Kinematics}
\label{kinematics}


 \begin{figure}    
 \centerline{\includegraphics[width=\textwidth,clip=]{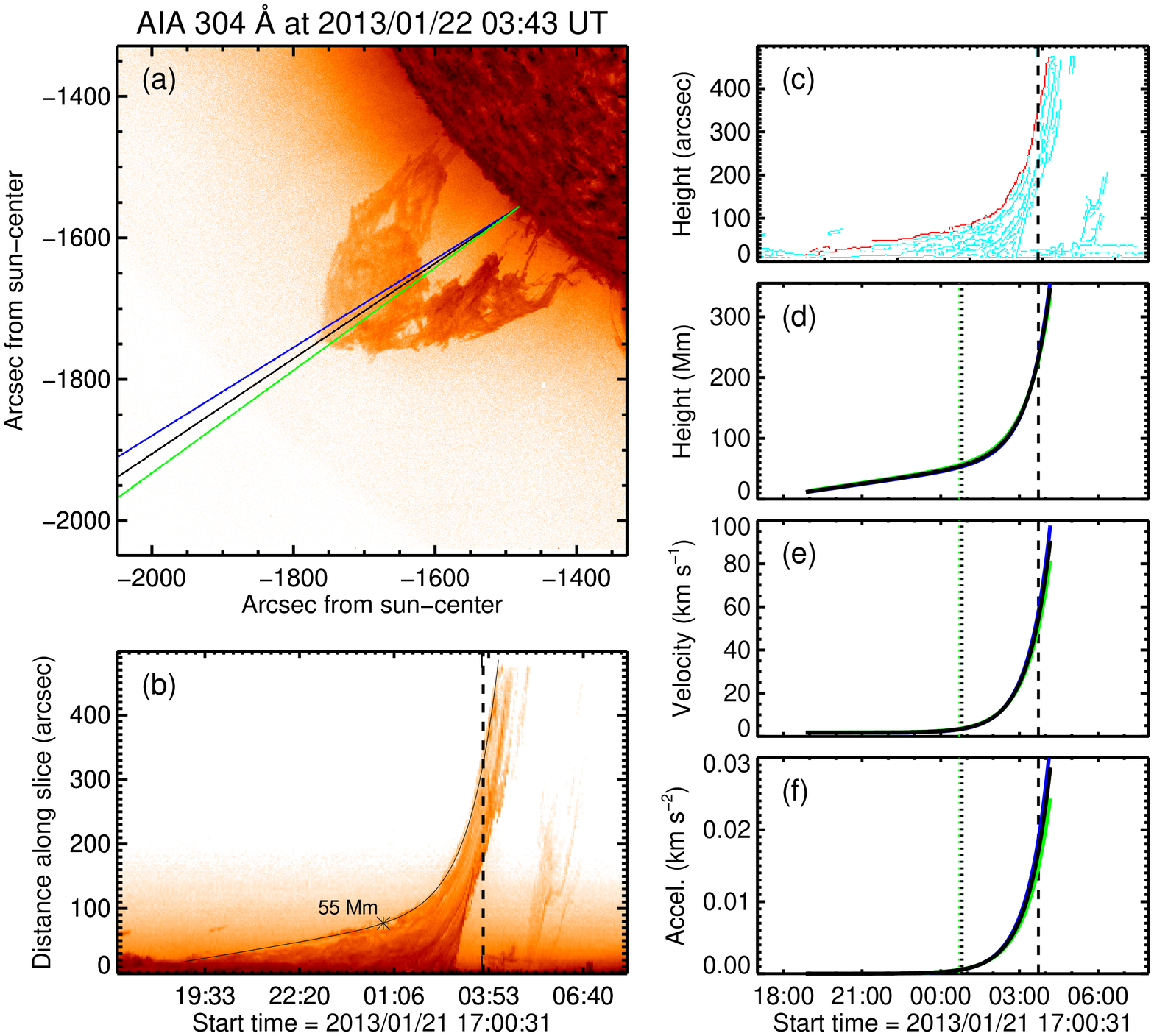}}
 \caption{An illustration of the kinematics procedure using event \href{http://aia.cfa.harvard.edu/filament/index.html?search=0620}{No. 0620}. 
 (a) Trajectories used for tracking. The black slice is selected manually, and the others are offset by 2$^{\circ}$ in either direction.
 (b) Height-time image for the black slice in Panel A. The dashed line indicates the time shown in Panel A, the 
 solid curve is the fit result, and the asterisk denotes the fast-rise onset point. 
 (c) Output of the Canny edge detection algorithm applied to Panel B. The red pixels are used as individual height measurements.  
 (d) Fit to the red pixels in Panel C. The dotted lines indicate the onset of the fast-rise phase, and the colors correspond to results from the different slices in Panel A.
 (e) Velocity and (f) acceleration profiles for the height profile in Panel D. 
 A movie showing the evolution of the eruption along the main slice is provided 
 on the catalog website. 
 }
 \label{f-procedure}
 \end{figure}

Filament eruptions often exhibit distinct slow- and fast-rise phases, which 
has been explored in a number of 
case studies (e.g. \citealt{Sterling04, Sterling05, Chifor06, Sterling07, Schrijver08, Sterling11, Joshi11, Koleva12}).
The slow portion is characterized by speeds of a few \kms{} or less that may persist 
for several hours or minutes for 
quiescent and active region filaments, respectively. 
Rapid acceleration to 
hundreds of \kms{} follows during the fast-rise phase, which may produce a CME 
or a confined eruption. The slow-rise phase is often attributed to 
gradual flux cancellation or emergence that builds and elevates the filament until 
it succumbs to an MHD instability or a fast-reconnection process that initiates  
the fast-rise phase \citep{Sterling07b, Sterling11, Fan15}. 
We will consider the underlying mechanisms in greater detail  
with respect to our results in \S\ref{discussion}. 
While previous low-corona studies have primarily  
relied on close inspection of individual events or a small ensemble, we examine the kinematics 
of 106 eruptions in our catalog to determine general properties of the slow- 
and fast-rise phases. 
In particular, we consider the transition between the two 
stages, using an analytic approximation to determine the onset of the fast-rise phase. 


\subsection{Procedure}
\label{procedure}

Eruptions on the limb were chosen for their heightened contrast and limited 
projection effects compared to those on the disk. Bulk motions are inferred by tracing 
the leading edges in 304 \AA{} observations and fitting  
curves to the resultant height-time profiles. Figure~\ref{f-procedure} illustrates this process 
for a single event.
To begin, we select the linear slice 
that best characterizes the overall trajectory of the filament (Panel A). 
Additional slices offset by two degrees in either direction 
are identically processed for error estimation.
Emission along these lines is binned to 300 pixels  
and interpolated onto a uniform grid, yielding a spatial 
resolution of $\sim$1.5'', which varies slightly depending on the length of the slice, 
dictated by FOV position. 
The resultant 
light curves are stacked against subsequent observations 
to produce height-time images like the one shown in Panel B. 
To further improve signal-to-noise, averages of 10 frames were  
used for each column measurement. This averaging corresponds to an effective 
exposure time of 29 seconds over a period of 2 minutes. Shorter averages 
were used for fast eruptions where a 2-minute time resolution was insufficient.

The height-time images are then further processed to improve contrast, which 
is typically limited to multiplying each row by its height to boost signal 
far from the limb and thresholding the image above some multiple of its median value.  
Smoothing is applied for particularly noisy cases, and base differencing (background subtraction) 
is used for events where the contrast was especially low. 
The Canny edge detection algorithm \citep{Canny86} is then employed to extract  
leading edges from the height-time images. Panel C shows the 
application of the Canny algorithm to the image in Panel B. Pixels highlighted 
in red, which form the uppermost edge, are used as the individual height measurements. 
These points are extracted automatically once the edge detection has been applied, but 
their time range must be selected manually. This consideration is particularly important 
for the start time because it effectively defines the onset of the slow-rise phase. 


\subsection{Analytic Approximation}
\label{equation}

 \begin{figure}    
 \centerline{\includegraphics[width=\textwidth,clip=]{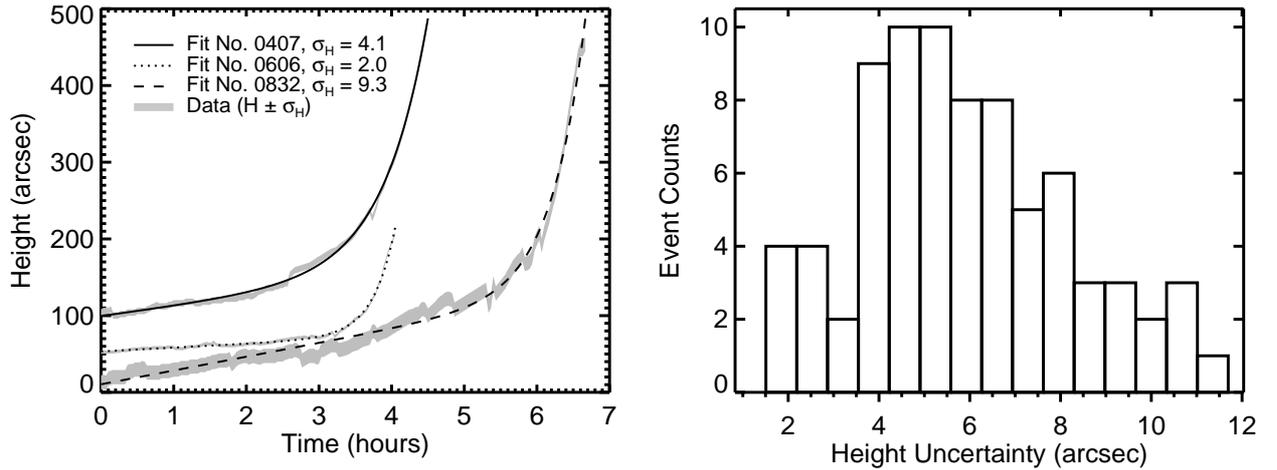}}
 \caption{\textit{Left}: Three examples with varying goodness-of-fits. 
 \textit{Right}: Histogram of height uncertainties, which 
 correspond to the values required for $\chi{}^{2}_{R} = 1.0$ for the fits to Equation~\ref{eq-fit}. }
 \label{f-uncertainties}
 \end{figure}

Several curves have been used in previous studies to fit the low-corona height 
evolution of eruptive prominences and CMEs in general. 
When there are distinct phases, the initial slow-rise profile is typically  
fit with a line, implying constant velocity (e.g. \citealt{Sterling05}), 
though at least one previous study has also employed a very slight 
constant-acceleration curve \citep{Joshi07}. 
The fast-rise phase and rapid acceleration 
that initiates it has been approximated by exponential (e.g. \citealt{Goff05, Williams05}), 
exponential-plus-initial-velocity \citep{Alexander02}, constant-acceleration 
(e.g. \citealt{Gilbert00, Kundu04}), and cubic \citep{Schrijver08} curves.
\citet{Cheng13} present an equation that combines a linear curve 
to treat the slow-rise phase and an exponential to treat the fast-rise 
after a time offset:

\begin{equation}  \label{eq-fit}
h(t) = c_0 e^{(t-t_0)/\tau} + c_1 (t - t_0) + c_2, 
\end{equation}

\noindent where $h(t)$ is height, $t$ is time, and $\tau$, $t_0$, $c_0$, $c_1$, and $c_2$ 
are free parameters. This model has the distinct advantage that both phases 
can be approximated by a single-function fit to the full dataset. 
It also provides a convenient method 
for determining the onset of the fast-rise phase, which can be 
defined as the point at which the exponential component of the velocity equals 
the linear (i.e. the total velocity is twice the initial), given by:

\begin{equation}  \label{eq-onset}
t_{\rm onset} = \tau \rm{ln}(c_1 \tau / c_0) + t_0
\end{equation}

\noindent We find this approximation to be best suited for our study, which aims 
to characterize the the slow- and fast-rise phases of many events in a uniform way. 
Fitting is accomplished using MPFIT, a non-linear least squares curve fitting 
package for IDL \citep{Markwardt09}. Panels D, E, and F of Figure~\ref{f-procedure} 
show the fit result and its time derivatives for our example event. 
Based on this approximation, we find that 
the initial slow-rise velocity is 1.8 $\pm{}$ 0.1 \kms{}, and the maximum velocity 
in the AIA  FOV is 91 $\pm{}$ 7 \kms{}. The onset of the fast rise phase occurs 
at 00:48 UT $\pm{}$ 5 minutes at a distance of 55 $\pm{}$ 5 Mm along the fit 
trajectory, which begins with an initial height of 0.015 \rsolar{} ($\sim$10 Mm).  
The onset point is therefore 65 Mm above the limb \textit{along the fit trajectory},  
corresponding to a radial height of 64 $\pm{}$ 5 Mm.  
At the onset 
of the fast-rise phase, the acceleration is 0.57 $\pm{}$ 0.03 \msq{}, and the final acceleration 
is 29 $\pm{}$ 4 \msq{}. Table~\ref{t-fit} lists the range of fit parameters 
found for the 78 events that can be satisfactorily described by this model. Details 
on the events that were not fit are given in \S\ref{results}.


\begin{table}
\caption{Equation~\ref{eq-fit} $h(t)$ fit variable statistics for 78 events. $t$ is in seconds 
relative to the start of observations, and $h$ is in arcsec  
along the trajectories shown in Figure~\ref{f-kin_summary}. 
}
\label{t-fit}
\begin{tabular}{l|cc|cc}
\hline
Variable & Min & Max & Mean $\pm{}$ SD & Median $\pm{}$ MAD \\
\hline
$\tau$ & 260 & 19000 & 2700 $\pm$ 2500 & 2100 $\pm$ 1400 \\
$t_0$ & 2600 & 130000 & 35000 $\pm$ 22000 & 33000 $\pm$ 16000 \\
$c_0$ & 37 & 1400 & 240 $\pm$ 200 & 180 $\pm$ 130 \\
$c_1$ & 0.00028 & 0.020 & 0.0029 $\pm$ 0.0027 & 0.0026 $\pm$ 0.0015 \\
$c_2$ & 30 & 310 & 120 $\pm$ 61 & 110 $\pm$ 49 \\
\hline
\end{tabular}
\end{table}

Two strategies are employed to quantify uncertainty, and errors quoted are the sum 
of both. The first is to identically process two adjacent slices offset by 
2$^{\circ{}}$ on either side of the original. On average, standard deviations from these results 
account for $\sim$65\% of the velocity, acceleration, and time uncertainties, but they may contribute 
as little as 30\% or as much as 90\% for a particular event. 
The second strategy is to perform 100 Monte Carlo (MC) simulations to estimate uncertainties 
from the fit parameters by randomly varying our height measurements within some assumed 
error and refitting Equation~\ref{eq-fit}. Since there are no standard errors for height measurements 
obtained by our edge detection method, 
the assumed height errors are chosen to yield a reduced 
chi-squared ($\chi{}^{2}_{R}$) of 1.0 for the fit. 
This uncertainty is therefore a proxy for the goodness of fit. 
In the example above, height errors of 5.2$''$ are required for $\chi{}^{2}_{R} = 1.0$, 
which is very close to the average value of 5.8$''$ across all events. 
5.8$''$ corresponds to $\sim$10 AIA pixels and around 4 pixels on the binned 
height-times images used to obtain the measurements.
These values account for 58\% of the fast-rise onset height errors on average, 
while uncertainties associated with the three separate slices 
and MC realizations typically account 
for 28\% and 14\% of the onset height uncertainties, respectively. 
The left panel of Figure~\ref{f-uncertainties} shows the data, errors, and associated fits for 
three examples with varying height uncertainties, and the right panel shows the 
total distribution of height uncertainties. 


\subsection{Solar Rotation}
\label{rotation}

As we will see in the next section, slow-rise velocities of $\lesssim$ 1 \kms{} are found 
for a number of events and have also been reported in the literature (e.g. \citealt{Sterling04, Isobe07, Liu12c,Regnier11,Reeves15}). 
It is important to note that solar rotation alone can produce apparent rise and fall 
velocities of this magnitude. Accounting for this effect requires an understanding of the 
projection geometry and the longitudinal extent of the filament, consideration of which is beyond 
the scope of this work. The effect is minor and should be at least 
partially washed out in our statistics, but we nevertheless include this 
discussion to demonstrate that very small velocities for specific events should be treated 
with some caution. 

\begin{figure}    
 \centerline{\includegraphics[width=\textwidth,clip=]{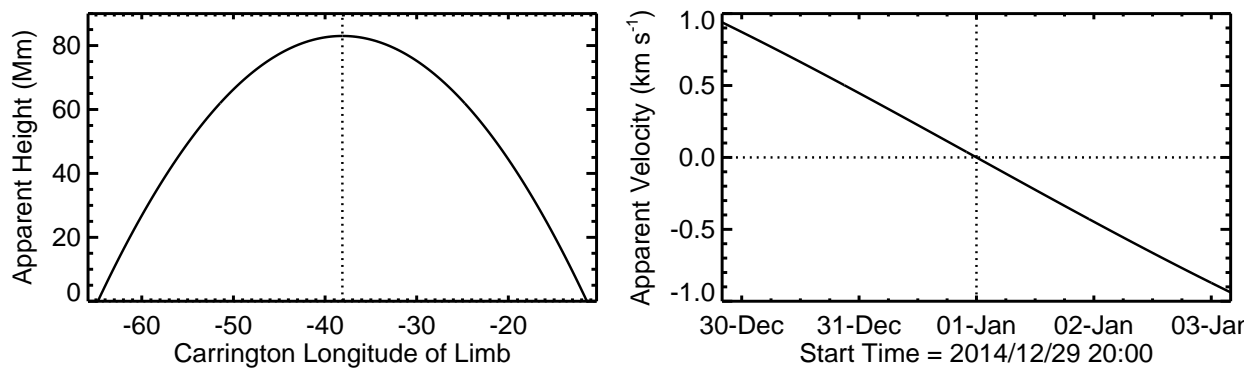}}
 \caption{The effect of solar rotation on apparent height (left) and velocity (right) 
 over a 4-day period for a prominence with 
 an actual height of 83 Mm, a latitude of 45$^{\circ{}}$, and a Carrington longitude of -38.1$^{\circ{}}$.
 }
 \label{f-rotation}
 \end{figure}

\citet{Foullon06} provide the following expression for the height of an arbitrarily thin 
prominence sheet above the limb, as seen from Earth:

\begin{equation} \label{eq-rotation}
\frac{h(L)}{\rsolar{}} = \left(\frac{h_0}{\rsolar{}} + 1\right)\cos{}(L - L_0) - 1,
\end{equation}

\noindent where $h(L)$ is apparent height, $h_0$ is actual height, $L$ is the Carrington 
longitude of the limb, and $L_0$ is the Carrington longitude of the prominence. 
Figure~\ref{f-rotation} illustrates this effect over a 4-day period 
for a hypothetical prominence with a 
height of 83 Mm (our average fast-rise onset height) at a latitude of 45$^{\circ{}}$ 
and a Carrington longitude ($L_0$) of -38.1$^{\circ{}}$, which corresponds to that  
of the east limb at 2014/01/01 00:00 UT. We see that  
rise and fall velocities approaching 1 \kms{} can be observed when a prominence 
begins to rotate onto and off of the limb. This could cause non-negligible over- or underestimations 
of the slow-rise velocites for the slower events in our sample, depending on their projection geometries
and longitudinal extents.

\subsection{Kinematics Results}
\label{results}


 \begin{figure}    
 \centerline{\includegraphics[width=0.7\textwidth,clip=]{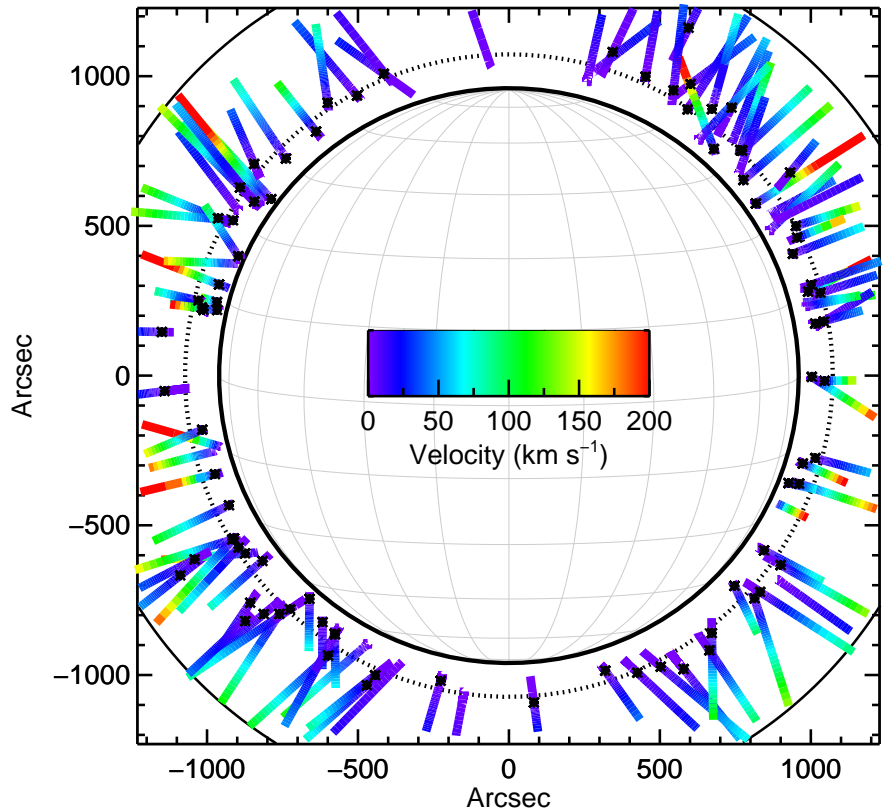}}
 \caption{Positions and velocity profiles from the kinematics study. The asterisks 
 denote the fast-rise phase onset point for events fit with Equation~\ref{eq-fit}. Events 
 without onset points are described by exponential functions. The dashed circle 
 represents the average radial onset height of 83 Mm, and the rounded corner edges 
 mark the boundary of the telescope's filter.
 }
 \label{f-kin_summary}
 \end{figure}
 

 \begin{figure}    
 \centerline{\includegraphics[width=\textwidth,clip=]{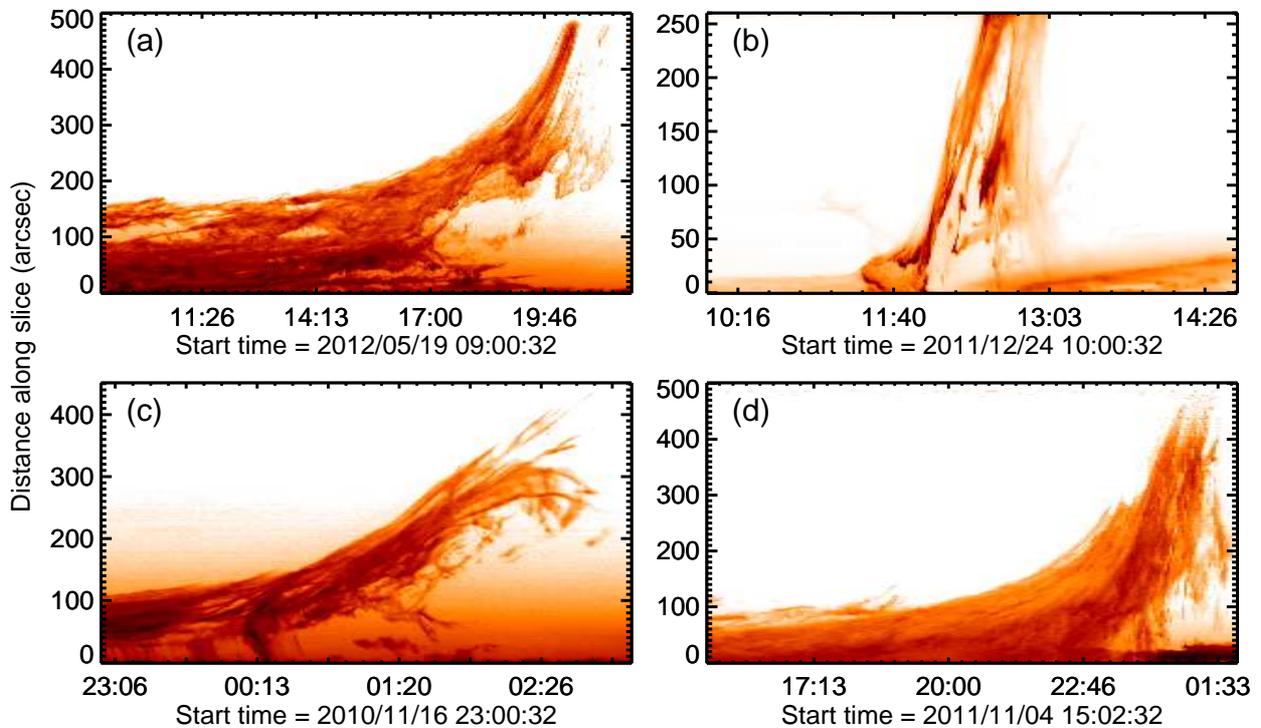}}
 \caption{Representative examples of the ``bad" events that could not be 
 described by Equation~\ref{eq-fit} and were excluded from further analyses.
 See \S\ref{results} text for details.
 }
 \label{f-bad}
 \end{figure}

Figure~\ref{f-kin_summary} summarizes the kinematics results, showing the 
positions and velocity profiles of each eruption. 
Of the 106 events, 78 (74\%) 
have distinct slow- and fast-rise phases that 
are well-fit by the 
linear-plus-exponential model proposed by 
\citet{Cheng13}.
These include 44 QS, 22 PC, 8 IP, and 4 AR eruptions. 
Representative examples of the remaining 
events that cannot be satisfactorily described
by Equation~\ref{eq-fit} are shown in Figure~\ref{f-bad}. 
Corresponding to the figure panels, these events 
cannot be fit because they a) most commonly do not exhibit 
slow-rise phases or have no marked transition between phases, b) have rise profiles 
too complicated to fit into the simple two-phase dichotomy, 
c) have distinct phases but the transition into the fast-rise is 
decidedly not exponential, or d) exhibit dynamics 
that make it difficult to 
infer bulk motions from that of their leading edges. 
These events include 15 QS, 9 PC, 2 IP, and 2 AR events, 
and they are represented in Figure~\ref{f-kin_summary} 
with exponential fits. 
While it would be interesting to consider these misfits in detail, they are 
excluded from all further analysis and left for future work. 


\begin{table}
\caption{Kinematics Results.
}
\label{t-lognormal}
\begin{tabular}{l|cc|cc|cc}
\hline
Parameter & Min & Max & Mean $\pm{}$ SD & Median $\pm{}$ MAD & ($\mu{} \pm{} \sigma{}$)\tabnote{Lognormal fit parameters.} & ($A^2$)\tabnote{Anderson-Darling goodness-of-fit statistic; low values imply better fits.} \\
\hline
$h_{\rm{onset}}$\tabnote{Radial fast-rise onset height.}~~(Mm) & 27 & 237 & 83 $\pm$ 41 & 71 $\pm$ 32 & 4.3 $\pm$ 0.51 & 0.35 \\
$\Delta{}h_{\rm{onset}}$\tabnote{Slow-rise phase displacement along fit trajectory (relative onset height).}~~(Mm) & 2.6 & 89 & 29 $\pm$ 22 & 22 $\pm$ 17 & 3.1 $\pm$ 0.86 & 0.37 \\
$\Delta{}t_{\rm{slow-rise}}$ (hrs) & 0.36 & 23 & 4.4 $\pm$ 3.7 & 3.2 $\pm$ 2.7 & 1.1 $\pm$ 0.93 & 0.37 \\
$v_{\rm{slow-rise}}$ (\kms{})  & 0.21 & 16 & 2.1 $\pm$ 2.1 & 1.9 $\pm$ 1.1 & 0.50 $\pm$ 0.67 & 0.81 \\
$v_{\rm{fov-max}}$ (\kms{}) & 6.0 & 417 & 106 $\pm$ 85 & 82 $\pm$ 60 & 4.4 $\pm$ 0.83 & 0.18 \\
$v_{\rm{500''}}$\tabnote{Parameters at 500$''$ above the limb; excludes 6 confined events.}~~(\kms{}) & 13 & 753 & 183 $\pm$ 169 & 126 $\pm$ 110 & 4.8 $\pm$ 0.84 & 0.38 \\
$a_{\rm{onset}}$ (\msq{}) & 0.032 & 29 & 1.9 $\pm$ 3.7 & 0.82 $\pm$ 1.5 & -0.28 $\pm$ 1.1 & 0.53 \\
$a_{\rm{fov-max}}$ (\msq{}) & 0.29 & 970 & 111 $\pm$ 183 & 40 $\pm$ 96 & 3.6 $\pm$ 1.5 & 0.38 \\
$a_{\rm{500''}}$$^e$ (\msq{}) & 0.64 & 1760 & 199 $\pm$ 362 & 57 $\pm$ 175 & 4.0 $\pm$ 1.6 & 0.43 \\
$n_{\rm{onset}}$\tabnote{Decay index at the fast-rise onset point; see Section 5.1.} & 0.69 & 2.0 & 1.1 $\pm$ 0.28 & 1.1 $\pm$ 0.22 & 0.089 $\pm$ 0.25 & 0.56 \\
\hline
\end{tabular}
\end{table}

It is also clear that the kinematics study is biased toward QS and PC eruptions from the 
proportion of AR and IP events in this sample compared to the catalog. 
This selection effect is primarily due to the 
semi-automated method described in \S\ref{procedure}, which requires that 
the trajectory be free of additional structures that cannot be easily subtracted away. 
The more dynamic environments of AR and IP filaments thus precluded 
a clean isolation of the eruption front by our edge detection method in a 
number of additional events that we attempted to process. 
This effect is exacerbated by the smaller size and lower initial heights of AR 
filaments because they consequently spend less time above the limb. Several 
otherwise suitable events were also excluded because a significant 
portion of their slow-rise phases occurred behind the limb or on the disk. 
For this reason, our AR sample includes somewhat larger filaments 
than average for that type. 
This bias combined with the small number 
of events means that our AR results may not accurately represent 
the general population, and additional work is therefore needed to obtain 
robust statistics.
Refinements to our technique can likely improve its applicability, but that 
is beyond the scope of this work. 

Figures~\ref{f-onsets},~\ref{f-velocity}, and~\ref{f-acceleration} show distributions of the
fast-rise onset heights, slow-rise durations, velocities, and accelerations
for the 78 events fit with Equation~\ref{eq-fit}. Cumulative distributions are 
also plotted for QS and PC events, along with a single curve for 
AR and IP eruptions, which are combined due to their small sample sizes.  
The histograms are best described by lognormal distributions, reminiscent  
of CME speeds \citep{Yurchyshyn05,Zhang06, Bein11}, 
and lognormal fits are plotted in red. An independent variable, $x$, is 
said to be lognormally-distributed when ln($x$) has a normal distribution.
Note that Figure~\ref{f-acceleration} plots the natural 
logarithm of acceleration because of the particularly large ranges
and therefore exhibits a normal distribution.
Physically, lognormal distributions may arise when a quantity is the 
multiplicative product of several independent variables, suggesting 
a multiplication of several independent processes, as opposed 
to the sum of independent variables for a normal distribution. 


 \begin{table}
\caption{Mean $\pm$ SD Kinematics Results by Type.
}
\label{t-kin_types}
\begin{tabular}{l|c|cccc}
\hline
 & \multicolumn{5}{c}{Type} \\
Parameter & All & AR & IP & QS & PC \\
\hline
Event Count & 78 & 4 & 8 & 44 & 22 \\
\hline
$h_{\rm{onset}}$ (Mm) & 83 $\pm$ 41 & 48 $\pm$ 16 & 60 $\pm$ 24 & 82 $\pm$ 42 & 101 $\pm$ 38 \\
$\Delta{}h_{\rm{onset}}$ (Mm) & 29 $\pm$ 22 & 13 $\pm$ 11 & 20 $\pm$ 9.4 & 25 $\pm$ 19 & 43 $\pm$ 24 \\
$\Delta{}t_{\rm{slow-rise}}$ (hrs) & 4.4 $\pm$ 3.7 & 1.2 $\pm$ 0.82 & 2.2 $\pm$ 1.9 & 3.7 $\pm$ 2.8 & 7.2 $\pm$ 4.4 \\
$v_{\rm{slow-rise}}$ (\kms{})  & 2.1 $\pm$ 2.1 & 2.3 $\pm$ 0.41 & 4.1 $\pm$ 4.8 & 1.9 $\pm$ 1.5 & 1.7 $\pm$ 0.89 \\
$v_{\rm{fov-max}}$ (\kms{}) & 106 $\pm$ 85 & 158 $\pm$ 95 & 218 $\pm$ 111 & 108 $\pm$ 74 & 53 $\pm$ 33 \\
$v_{\rm{500''}}$ (\kms{}) & 183 $\pm$ 169 & 359 $\pm$ 180 & 390 $\pm$ 216 & 183 $\pm$ 146 & 80 $\pm$ 52 \\
$a_{\rm{onset}}$ (\msq{}) & 1.9 $\pm$ 3.7 & 4.1 $\pm$ 3.4 & 6.6 $\pm$ 9.2 & 1.4 $\pm$ 1.6 & 0.60 $\pm$ 0.49 \\
$a_{\rm{fov-max}}$ (\msq{}) & 111 $\pm$ 183 & 320 $\pm$ 261 & 336 $\pm$ 316 & 97 $\pm$ 132 & 21 $\pm$ 24 \\
$a_{\rm{500''}}$ (\msq{}) & 199 $\pm$ 362 & 505 $\pm$ 402 & 628 $\pm$ 589 & 176 $\pm$ 305 & 33 $\pm$ 47 \\
$n_{\rm{onset}}$ & 1.1 $\pm$ 0.28 & 1.0 $\pm$ 0.18 & 1.1 $\pm$ 0.21 & 1.2 $\pm$ 0.29 & 1.1 $\pm$ 0.28 \\
\hline
\end{tabular}
\end{table}


The lognormal probability, $f(x)$, and cumulative 
probability, $F(x)$, distributions are given by:

\begin{equation} \label{eq-lognormal}
f(x) = \frac{1}{\sqrt{2\pi}\sigma{}x}\textrm{exp} \left(-\frac{(\ln{x} - \mu)^2}{2\sigma{}^2} \right); 
~~F(x) = \frac{1}{2}+\frac{1}{2}\textrm{erf} \left(\frac{\ln{x} - \mu}{\sqrt{2}\sigma}\right),
\end{equation}

\noindent where $\mu$ is the mean and $\sigma$ is the standard deviation of ln($x$).
A least squares fit to the empirical cumulative probability 
distribution is used to approximate $F(x)$, avoiding the unnecessary binning that would be needed to fit $f(x)$. 
We employ the modified Anderson-Darling ($A^2$) test to quantify the goodness-of-fit, 
which tells us if the empirical distribution is \textit{unlikely} to be drawn 
from the fitted distribution. 
The null hypothesis for this test is that the empirical distribution \textit{is} drawn from 
the fitted distribution, which can be rejected at the 0.05 significance level if 
$A^2 > 0.752$ and at the 0.25 level if $A^2 > 0.470$ \citep{Stephens86}. 
Low values of $A^2$ therefore imply better fits. 
 Table~\ref{t-lognormal}
 lists several statistical measures for each of the total distributions, including 
 the lognormal fit parameters and $A^2$ values, and Table~\ref{t-kin_types} lists the arithmetic 
 means and standard deviations for each quantity by filament type. 
With reference to \citet{Bein11} and \citet{Limpert01}, we 
use $\mu{}^* = e^{\mu}$ and $\sigma{}^* = e^{\sigma}$ to refer to the median 
and multiplicative standard deviation of the fitted distribution, respectively. 
The 68\% confidence interval for a lognormal distribution then spans from 
 $\mu{}^* \div{} \sigma{}^*$ to $\mu{}^* \times{} \sigma{}^*$.


\subsubsection{Fast-Rise Phase Onset Heights \& Slow-Rise Durations}
\label{onset}

\begin{figure}    
 \centerline{\includegraphics[width=\textwidth,clip=]{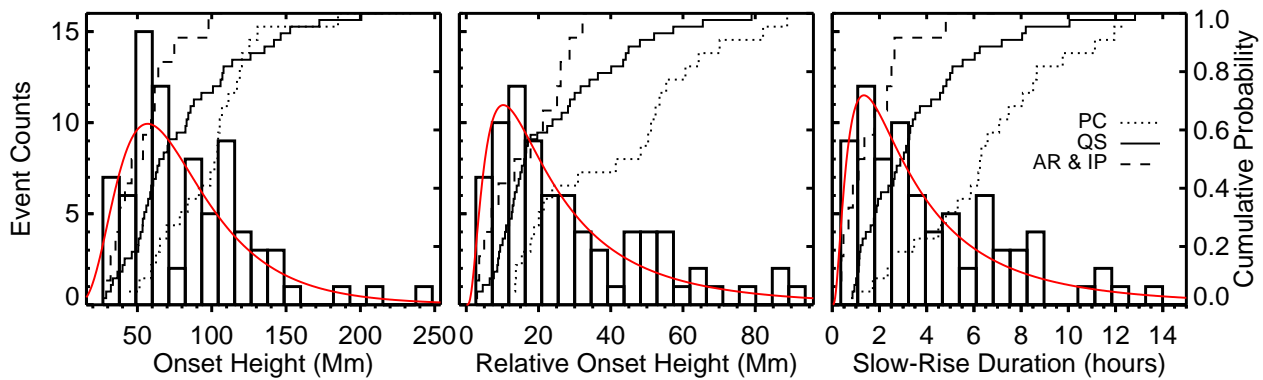}}
 \caption{\textit{Left}: Radial fast-rise onset heights. 
 \textit{Center}: Slow-rise phase displacements along the fit trajectories. 
  \textit{Right}: Slow-rise phase durations. 
  The histograms show the overall count distributions (left axes), 
  the black curves show the cumulative probability distributions for specific types (right axes), 
 and the red curves show lognormal fits to the histograms. 
 }
 \label{f-onsets}
 \end{figure}

The left panel of Figure~\ref{f-onsets} shows the distribution of radial 
fast-rise onset heights ($h_{\rm onset}$) derived from the prominence positions 
at  $t_{\rm onset}$ (Equation~\ref{eq-onset}).
We find mean and median values of 83 and 71 Mm, respectively, and the fitted lognormal distribution has 
a median ($\mu{}^*$) of 74 Mm and a  68\% confidence interval that spans from 44 to 122 Mm. 
Compared to 
QS eruptions, for which $\overline{h}_{\rm{onset}}$ = 82 Mm, AR and IP 
filaments tend to reach their critical points at $0.59\times$ and $0.74\times$ lower heights, 
respectively, while PC events transition into the fast-rise phase at $1.24\times$ 
larger altitudes. 
We can also consider the distances traveled along the 
fit trajectories, which may be non-radial (see Figure~\ref{f-kin_summary}), by subtracting 
the initial leading edge positions from their positions  
at $t_{\rm onset}$. 
This relative onset height ($\Delta{}h_{\rm{onset}}$) reflects the magnitude 
of the slow-rise phase displacement, and its distribution is shown 
in the middle panel of Figure~\ref{f-onsets}. 
$\mu^*$ = 21.3 Mm for $\Delta{}h_{\rm{onset}}$, very close to the actual median value 
of 21.8 Mm, and the 68\% confidence interval ranges from 9 to 50 Mm. The 
relative onset heights are $0.51\times$ lower, $0.80\times$ lower, and $1.75\times$ higher for AR, IP, 
and PC events, respectively, compared to the QS average of 25 Mm. Some 
of the differences between both the radial and relative onset heights across 
filament types can be attributed to a positive correlation between latitude 
and onset height found for the entire sample, which will be 
discussed in \S\ref{latitude} with respect to the decay index of the magnetic field. 

\citet{Liu12} also examine the fast-rise onset height, which they refer to 
as the critical height, using STEREO observations of 362 prominences that were
detected and tracked by an automated system \citep{Wang10}. 
Height-time measurements from the fast-rise phase were fit with a line, 
and the critical height was defined as the intersection between this line 
and the height profile prior to the fast-rise phase. This method would 
give somewhat larger heights if applied here, since our onset heights 
refer to the base of an exponential growth curve (Equation~\ref{eq-onset}), 
but this difference is very small compared to the range of heights found.  
\citet{Liu12} find an average critical height of 77 Mm (0.11\rsolar{}), 
which is 6 Mm less than our average but still well within 
one standard deviation. 



The right panel of Figure~\ref{f-onsets} shows the distribution of slow-rise 
phase durations, or the time required to cover $\Delta{}h_{\rm{onset}}$. 
The average across all types is 4.4 hours, and the type averages 
are sorted as expected, with AR and PC events at the low and high ends. 
The 68\% confidence interval spans 1.2 to 8.0 hours, which covers 
most of the values reported in the literature (e.g. \citealt{Sterling05, Joshi07,Reeves15}). 
The minimum slow-rise phases were 22 and 24-minutes for an IP 
and AR event, respectively. The short end found in the literature 
is around 10 minutes, reported by \citet{Sterling05} for an AR eruption, 
though the authors note that the filament appeared to begin rising 
at least 10 additional minutes earlier. \citet{Chifor06} report 
and AR filament slow-rise duration of 34 minutes for a separate event. Our sample of AR events is quite small 
for reasons described above, which likely contributes to the paucity of 
brief slow-rise phases and biases our lognormal fit toward longer durations. 
AR eruptions may also exhibit only exponential growth (i.e. have no 
slow-rise phase), such as the two AR events that could not be fit by 
Equation~\ref{eq-fit} in our initial sample and the event described by \citet{Williams05}. 
In these cases, the slow-rise phase may be non-existent or may simply be indiscernible 
by the analysis technique or instrument resolution. At the other end of the spectrum, 
the longest slow-rise phase is 23 hours for a PC eruption. This event is an extreme outlier 
at 3 hours outside of the 95\% confidence interval upper bound ($\mu{}^* \times{} \sigma{}^{*2}$) 
for the lognormal fit, already expected to be biased toward long durations, and indeed 
we could find no counterparts in the literature. 
\citet{Liu12c} report a slow-rise 
phase duration of $\sim$10 hours, which is one of the largest values previously reported and is 
much closer to the longer slow-rise phases we observe, excluding the outlier.


\subsubsection{Velocity \& Acceleration Distributions}
\label{velocity}

    \begin{figure}    
 \centerline{\includegraphics[width=\textwidth,clip=]{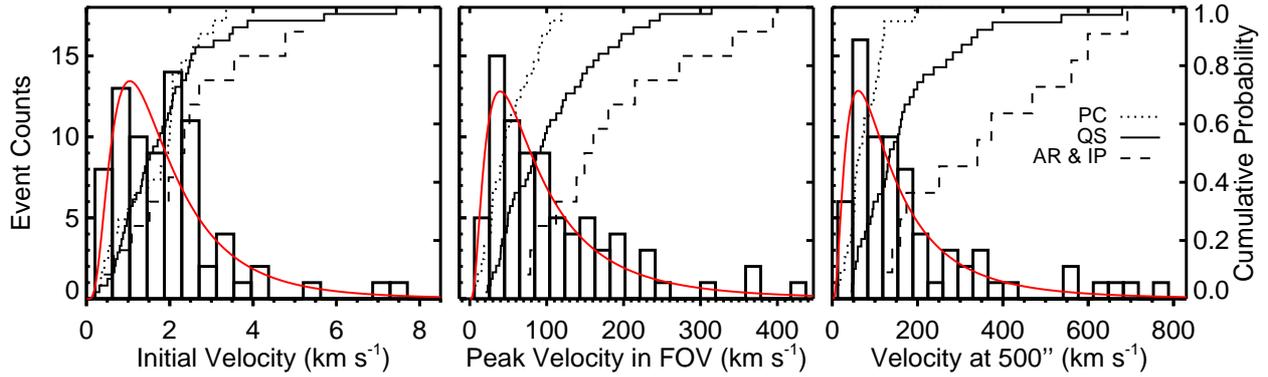}}
 \caption{\textit{Left}: Slow-rise velocity. 
 \textit{Center}: Peak velocity in FOV. 
  \textit{Right}: Velocity at 500'' above the limb (projected). 
The histograms, curves, and axes are as in Fig~\ref{f-onsets}.
  }
 \label{f-velocity}
 \end{figure}

\begin{figure}    
 \centerline{\includegraphics[width=\textwidth,clip=]{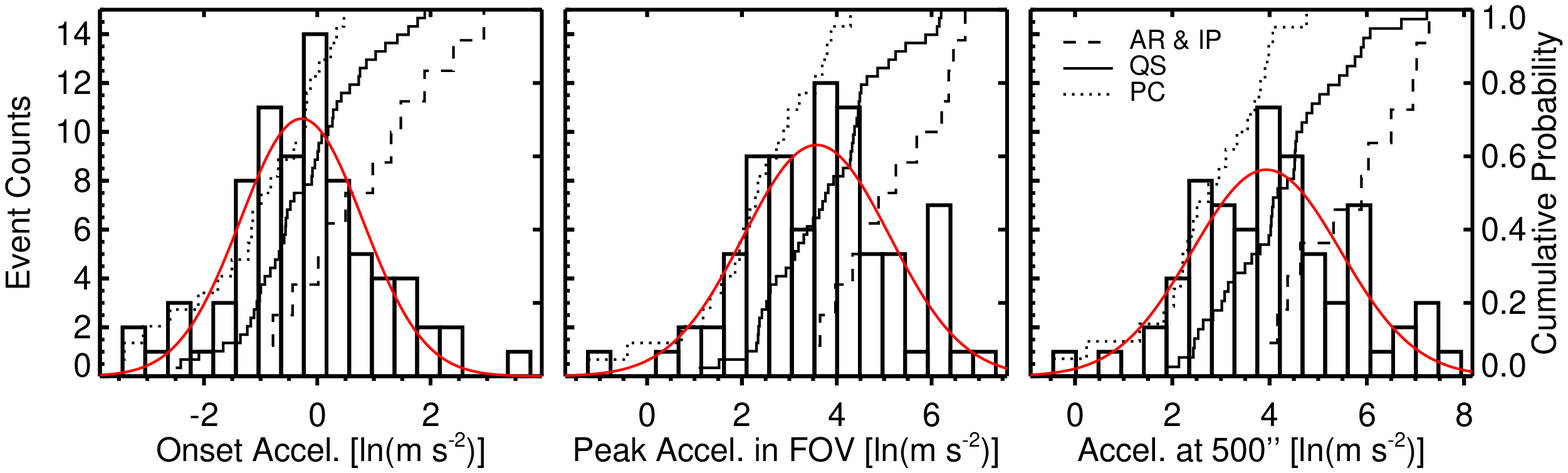}}
 \caption{\textit{Left}: Fast-rise onset acceleration. 
 \textit{Center}: Peak acceleration in FOV. 
  \textit{Right}: Acceleration at 500'' above the limb (projected). 
The histograms, curves, and axes are as in Fig~\ref{f-onsets}.
  }
 \label{f-acceleration}
 \end{figure}

Moving to the velocity distributions plotted in Figure~\ref{f-velocity}, we find that the initial 
slow-rise speeds range from a minimum of 0.2 to a maximum of 16 \kms{}. 
Recalling \S\ref{equation}, these values are derived from the linear 
component of Equation~\ref{eq-fit} and dominate the total velocity until the 
exponential component of the curve contributes  
equally at $t_{\rm onset}$ (Equation~\ref{eq-onset}). The lognormal fit 
to the $v_{\rm slow-rise}$ distribution is particularly bad, with 
$A^2 > 0.752$, meaning that we can reject the hypothesis that the 
velocities are lognormally-distributed with a $p$-value $<$ 0.05. 
Instead, the histogram is double-peaked with enhanced 
slow-rise velocities around  
1 and 2 \kms{}. 
However, if we consider only QS events, the values follow a lognormal distribution 
very closely to $A^2$ = 0.25. (In fact, all of the QS-only distributions 
except for $v_{\rm fov-max}$ are somewhat better-fit by 
lognormal profiles than the corresponding total distributions, with $\overline{A^2} = 0.31$ vs. 0.42. 
While we do not show the type-specific histograms and fit profiles, the 
characters of their distributions may be gleaned from the cumulative 
distributions in Figures~\ref{f-onsets}--\ref{f-acceleration}.) 

The total $v_{\rm slow-rise}$ distribution deviates from log-normality primarily 
because the PC-only distribution, which has 
a large enough sample to draw meaningful conclusions from, 
is \textit{not} lognormal to a significance of over 99\% ($A^2 = 1.1$). Instead, it is better 
fit by a single Gaussian with a mean of 1.7 and a variance of 1 \kms{} ($A^2 = 0.55$). The low  
number of events in the AR and IP distributions preclude us from concluding anything about their 
general $v_{\rm slow-rise}$ distributions, but the limited information hints at  
log-normality for IP and normality for AR eruptions. In light of this, the other 
panels of Figure~\ref{f-velocity} reveal an intriguing development. 
The middle panel shows the maximum velocities attained in the AIA FOV, 
and the right panel shows the fit velocity at a fixed height of 500$''$ above the limb, 
which is roughly the maximum height that AIA can observe for structures 
near the FOV corners. 
These latter velocities are included because the maximum observable height, where 
$v_{\rm fov-max}$ is generally taken, varies depending on FOV position (see Figure~\ref{f-kin_summary}). 
Low-latitude fits are therefore projected for the 500$''$ comparison, and 
6 confined events are excluded because the filament material does not reach that height. 

Both of these velocity distributions exhibit robust lognormal profiles despite 
having emerged from a slow-rise distribution that is significantly less so. 
This is true even for the PC-only distributions ($A^2 \approx 0.33$). 
Lognormal profiles are also found for each of the acceleration distributions 
shown in Figure~\ref{f-acceleration}, which is particularly significant for the 
fast-rise onset accelerations ($a_{\rm onset}$) because it indicates 
that the velocity distribution shift begins immediately 
at the start of fast-rise phase. 
We might draw two inferences from this. 
First, the normal distributions 
of PC and perhaps AR slow-rise velocities hint at a more uniform 
process than the IP and QS counterparts. This finding may be understood in terms 
of the more varied environments and size-scales of IP and QS filaments, 
which provide a more diverse set of influences to impact the slow-rise pace. 
Second, the lognormal distributions for the fast-rise velocities of all types 
hint at the multiplicative contribution of several processes. This result may 
be understood in terms of the various, and not necessarily mutually exclusive, 
fast-rise onset mechanisms (to be discussed further in \S\ref{discussion})
in combination with environmental influences. We also see that the
lognormal distribution of CME speeds reported in previous works 
emerges early in the height evolution of prominence eruptions. 

Now we turn briefly to how our velocity and acceleration values 
differ by filament type. As 
with the CME speeds in \S\ref{cmes}, we find that IP eruptions 
are generally the fastest of the four types. The average 
IP slow-rise velocity is 4.1 \kms{}, which is about 1.8$\times$, 2.2$\times$, 
and 2.4$\times$ faster than the AR, QS, and PC averages, respectively. 
Similar relationships between types are found for the other velocity and acceleration quantities. 
Accelerations at $t_{\rm onset}$ range from a minimum of 0.03 \msq{} for a PC eruption to a maximum 
of 29 \msq{} for an IP event, with 
an overall average value of 1.9 and a 68\% confidence interval of 
0.25 to 2.3 \msq{}. After the fast-rise onset, the velocities 
grow exponentially to reach values at 500$''$ above the limb of between 
13 \kms{} for the slowest PC eruption and 750 \kms{} 
for the fastest IP event. 
The median value upon exiting the AIA FOV is 82 \kms{}. 
See Table~\ref{t-lognormal} and Table~\ref{t-kin_types} for 
the full list of statistical parameters. 
These values are broadly consistent with those reported 
in the literature, much of which has already been referenced. 


\subsection{Online Catalog Kinematics Content}
\label{kinematics_access}

The online catalog includes 
a version of Figure~\ref{f-procedure} and the associated movie 
for each of the 106 events 
in the kinematics study. Text files files 
containing the height vs. time data, 
fit parameters, and essential results are also provided. 
Information on finding and using these files is provided on the 
catalog website. 

 
\section{Discussion} %
\label{discussion} %


 \begin{figure}    
 \centerline{\includegraphics[width=\textwidth,clip=]{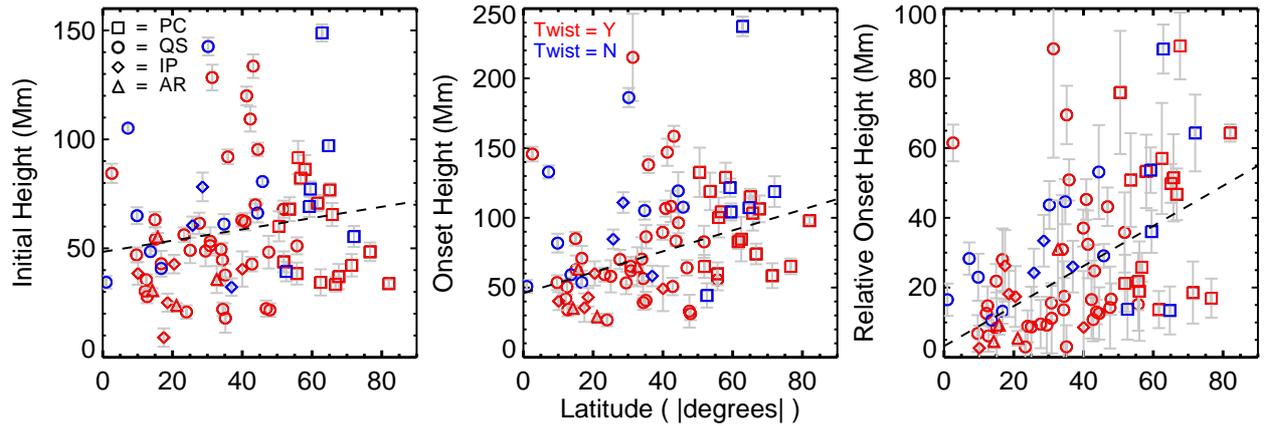}}
 \caption{Latitude versus initial height (left), onset height ($h_{\rm onset}$, middle), 
 and relative onset height ($\Delta{}h_{\rm onset}$, right), where the latitudes are those of the 
 onset points  
 in Figure~\ref{f-kin_summary}.
  }
 \label{f-latitude}
 \end{figure}

A few interesting patterns emerge from our results that are relevant to 
the broader discussion on filament eruption triggering. 
Initiation mechanisms for the fast-rise, rapid 
acceleration phase (or for the entire eruption if there is no slow-rise) 
are generally divided between ideal 
MHD instabilities and fast magnetic reconnection processes 
(see review by \citealt{Fan15}). The 
former includes the helical kink instability, discussed in \S\ref{kink}, 
and the torus instability, which occurs when a toroidal flux rope rises 
to a critical height above which it cannot be contained by 
the overlying magnetic field \citep{Kliem06}. 
Considering reconnection triggers, flux emergence \citep{Chen00} or cancellation \citep{Ballegooijen89} 
may lead to reconnection that stimulates the eruption, or breakout reconnection in 
the overlying field may prompt an eruption \citep{Antiochos99}. These 
mechanisms are not necessarily mutually exclusive, and reconnection may also   
facilitate an eruption triggered by the catastrophic loss of MHD equilibrum \citep{Lin00}.
Several mechanisms have also been proposed for the slow-rise phase, 
perhaps the most common of which is flux cancellation 
that increases the twist of a filament, slowly buoying it  
until an MHD instability or reconnection process 
triggers rapid acceleration (e.g. \citealt{Amari10,Aulanier10}). 
Alternatively, \citet{Reeves10} simulate the initial slow rise 
of a flux rope CME by means of resistive diffusion, and 
\citet{Fan07} discuss the possibility of a flux rope emerging 
from beneath the photosphere. 

\subsection{Onset Height, Latitude, \& Decay Index}
\label{latitude}

Figure~\ref{f-latitude} plots the initial prominence height, onset height,  
and relative onset height versus latitude, 
where the latitudes are those of the onset points in Figure~\ref{f-kin_summary}. 
Recalling \S\ref{onset}, the onset height ($h_{\rm onset}$) is a 
radial height above the limb, while the 
relative onset height ($\Delta{}h_{\rm onset}$) refers to a 
displacement along a particular fit trajectory. 
 We find a positive correlation between 
latitude and onset height, with a Pearson's correlation coefficient ($r$) 
of 0.30 and a $p$-value of 0.008. The relationship strengthens if we instead consider 
the relative onset height ($r$ = 0.43, $p$ = 7\e{-5}) or 
slow-rise phase duration ($r$ = 0.50, $p$ = 3\e{-6}). This effect is distinctly 
global in that if we divide the data into latitude bins of 
20$^\circ$, for instance, the correlation 
within a given bin is insignificant or non-existent. 
It is also distinct from a relationship between latitude and the initial prominence height, 
for which there is a weak and statistically insignificant 
positive correlation ($r$ = 0.14, $p$ = 0.13) that disappears if 
IP and AR events are excluded.


 \begin{figure}    
 \centerline{\includegraphics[width=\textwidth,clip=]{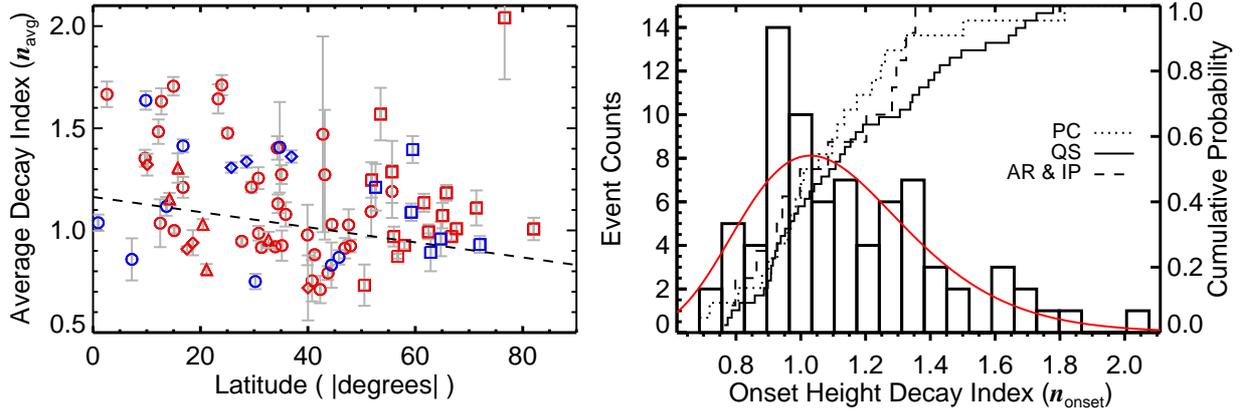}}
 \caption{
 \textit{Left:} Average decay index between 42 and 105 Mm vs. latitude. 
 Symbols and colors are as in Fig.~\ref{f-latitude}.
 \textit{Right:} Distributions of the decay indexes at the fast-rise 
 onset height. The histogram, curves, and axes 
 are as in Fig.~\ref{f-onsets}--\ref{f-acceleration}.
  }
 \label{f-decay}
 \end{figure}

The maximum height of non-eruptive prominences was suggested as 
a parameter of the magnetic field by \citet{Leroy84}. \citet{Makarov92}
expanded on this, using height variations among prominences of the 
same latitude as a proxy for longitudinal variations in the background  
field strength.  \citet{Makarov94} also reported that the height of polar 
crown prominences (latitudes greater than 55$^\circ{}$) decreased 
with increasing latitude using observations of many prominences over several solar cycles. 
We see this effect to some extent in our initial height 
data for the PC-only sample, which exhibit a minor negative correlation ($r$ = -0.21, $p$ = 0.35) between 
latitude and initial height. 
Upon erupting, the PC-only data shift to no correlation with latitude 
for onset height and a minor positive correlation ($r$ = 0.17, $p$ = 0.45) for   
relative onset height.

The critical height of a flux rope, above which 
no stable equilibria exist, can be related to a critical 
threshold in the vertical 
gradient of the magnetic field, or decay index \citep{Filippov00}: 

\begin{equation} \label{eq-decay}
n = - \frac{d(\ln{\textbf{B}})}{d(\ln{h})},
\end{equation}

\noindent where $\textbf{B}$ is the strapping field and $h$ is height above 
the photosphere.
This is the essential concept of the previously-mentioned torus instability, 
which refers specifically to toroidal (curved) flux ropes but may sometimes  
be used more generally to encompass all geometries.  
The critical decay index ($n_{\rm cr}$) for loss of confinement depends  
principally on the assumed flux rope configuration and is canonically 
1.0 for the straight \citep{van78} and 1.5 for the toroidal \citep{Bateman78} 
thin current channel approximations.
Both geometries represent limiting cases of the same physical process \citep{Demoulin10}.
Recent theoretical calculations and numerical simulations have 
found $n_{\rm cr}$ values between 1.0 and 2.0 under various  
assumptions that better approximate coronal conditions \citep{Kliem06,Fan07,Aulanier10,Demoulin10}. 
If we assume that many of the events in our sample are 
driven to erupt by this sort of mechanism, then a 
positive correlation between latitude and fast-rise 
onset height implies that the decay index is smaller 
at high latitudes for a given altitude.

To test this hypothesis, we estimate the decay index along 
our eruption trajectories using  
potential field source surface (PFSS) 
extrapolations of the coronal magnetic field 
based on line-of-sight magnetograms from the SDO's Helioseismic and Magnetic Imager (HMI, \citealt{Scherrer12}).
Decay index maps are obtained for the plane of the sky through Sun-center
using the SolarSoft IDL software package 
\href{http://www.hao.ucar.edu/FORWARD/}{FORWARD}\footnote{SolarSoft IDL FORWARD Package: \url{http://www.hao.ucar.edu/FORWARD/}} \citep{Gibson14},
which is capable of forward 
modeling various observables but in this case simply  
provides a convenient interface for SolarSoft's 
\href{http://www.lmsal.com/~derosa/pfsspack/}{PFSS package}\footnote{SolarSoft IDL PFSS Package: \url{http://www.lmsal.com/~derosa/pfsspack/}}.  
Details on the PFSS model are given by \citet{Schrijver03}.  
For each event, we record the decay index 
at the fast-rise onset height ($n_{\rm onset}$) and the average 
decay index between 42 and 105 Mm above the limb ($n_{\rm avg}$), the range of which is 
chosen for consistency with \citet{Liu08}. 
The left panel of Figure~\ref{f-decay} shows $n_{\rm avg}$ versus 
latitude, and we do find the expected negative correlation ($r$ = -0.22, $p$ = 0.06). 
This result, combined with the slight positive correlation between 
initial height and latitude, nicely explains the positive correlations 
between latitude and the related measures of fast-rise onset height, relative 
onset height, and slow-rise duration. 

The right panel of Figure~\ref{f-decay} shows a histogram 
of the fast-rise onset height decay indexes, and basic statistics 
are listed in Tables \ref{t-lognormal} and \ref{t-kin_types}. 
It is interesting to note that the $n_{\rm onset}$ 
distribution clusters around 1.0, closer to the straight flux rope approximation 
than the curved. 
The average value is 1.1 and is essentially the same for all four filament 
types, despite their differences in kinematics parameters, particularly 
onset height. This result is generally consistent with \citet{Filippov01} and \citet{Filippov13}, who 
also find critical decay indexes of around 1.0 for observed quiescent filament eruptions 
and therefore suggest that the forces associated with the axis of curvature 
are essentially unimportant for the equilibria of coronal flux ropes. 
However, we also find that a significant fraction (27\%) of our events have critical 
decay indexes greater than 1.3. Indeed, our results span the full range of aforementioned 
theoretical $n_{\rm cr}$ values derived from various flux rope and simulation configurations. 
We also find that 38\% of our events begin their fast-rise phases with decay 
indexes of less than 1.0 and 15\% do so with $n_{\rm onset} < 0.9$. 
This finding suggests that an appreciable fraction of the eruptions are initiated by other means, such 
as reconnection or the kink instability. 

Some limitations to our decay index estimates should be noted. 
First, we are looking at regions on the limb, where the PFSS extrapolations are 
constrained by observations a few days before or after the eruptions for the west 
and east limbs, respectively. 
Such time offsets are especially limiting for active regions, which are very likely 
to have evolved over those periods.  
Second, we use single 
values in the plane of the sky, which may not 
be perfectly aligned with a given filament.  
These considerations are particularly important for the discussion of eruption mechanisms 
because relatively small changes in the decay index can imply 
significant differences in the initiation process.
They might also explain two peculiarities in our decay index results. 
The curvature of AR region filaments, with their shorter lengths, is generally 
thought to be more important compared to that of their more extended QS counterparts, 
meaning that AR $n_{\rm cr}$ thresholds should be closer to the toroidal 
approximation of $\sim$1.5. 
We instead find an average $n_{\rm onset}$ of 1.0 for the 4 AR events in our sample, 
which might be due to the PFSS limitations. 
There is also a conspicuous outlier in Figure~\ref{f-decay}, with a high latitude 
and large decay index. This filament (\href{http://aia.cfa.harvard.edu/filament/index.html?search=0934}{No. 0934})
appears to be inclined such that the plane-of-sky decay index is not appropriate. 
More careful modeling and the use of STEREO observations to triangulate 
the filament positions would improve the decay index estimates.
We will discuss possibilities for additional work on 
this topic in \S\ref{future}.

\subsection{Kinematics \& Twist}
\label{speed_twist}

\begin{figure}    
 \centerline{\includegraphics[width=\textwidth,clip=]{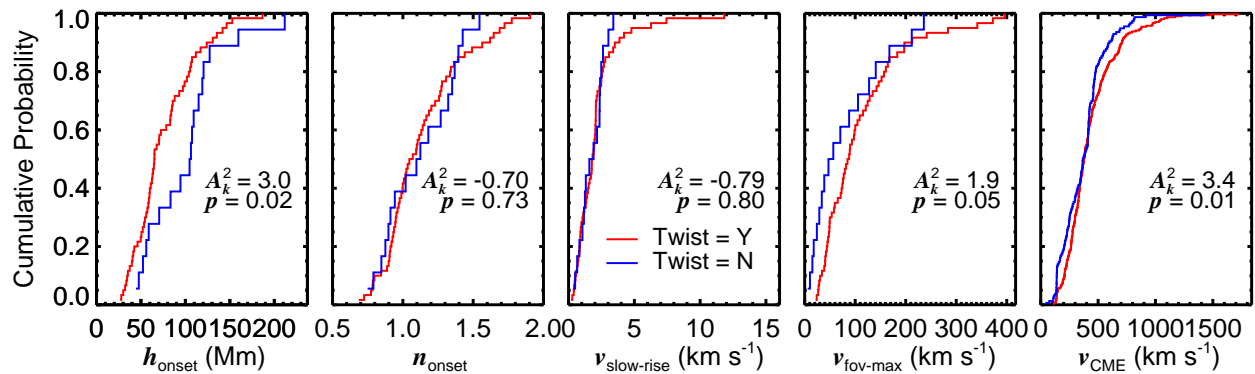}}
 \caption{Empirical cumulative probability distributions for events with and without 
 apparent twisting motions (see \S\ref{twist}). The right panel (CME speed) 
covers the entire catalog, while the others include the 78 events 
 in the kinematics study fit with Equation~\ref{eq-fit}. 
  }
 \label{f-twist_dist}
 \end{figure}

It is also interesting to consider possible relationships 
between kinematics and  
apparent twisting motions. 
Referring to \S\ref{twist}, ``apparent twist" may include 
motion that is induced during the eruption by surrounding 
features (i.e. the ``roll effect") or, more commonly, motion that is likely to be  
indicative of the intrinsic flux rope twist, such as 
untwisting motions observed in filament footpoints near the end 
of an event.  We hypothesize that on average, 
eruptions exhibiting twisting motions will have had greater 
helicity at their fast-rise onset points compared to those 
with no apparent twisting motions. 
Given this expectation, it is interesting to note that all but 2 of the 18 
events with no apparent twist lie very close to or above 
the fit lines in the middle and right of Figure~\ref{f-latitude} (latitude vs. onset-height). 
On average, events with no twist begin their 
fast-rise phases at 1.4$\times$ larger heights, 
and comparing the cumulative distributions in the left panel of Figure~\ref{f-twist_dist}
indicates a statistically significant difference ($A^2_k$ = 3.0 and $p$ = 0.02).
This result may be attributed to differences in the eruption 
mechanisms that preferentially affect more or less twisted 
filaments. One possibility is that highly-twisted flux ropes are more 
susceptible to the kink instability compared to less-twisted structures, which  
are more likely to succumb to the torus instability. These 
mechanisms should produce populations of lower and higher fast-rise 
onset heights, respectively, which would be superposed on a scatter of 
reconnection-driven events. The right panel of Figure~\ref{f-latitude}, in particular, 
exhibits a pattern somewhat similar to this. 


\citet{Fan07} find from MHD simulations that the 
field decreases with height more slowly for the kink instability 
compared to the torus instability. 
We might therefore expect to see differences in the distributions of fast-rise onset height 
decay indexes ($n_{\rm onset}$) between our two twist groups if there are indeed systematic 
differences in their initiation mechanisms. 
The second panel of Figure~\ref{f-twist_dist} shows that 
there is no such difference, and the same is true if 
we instead consider the average decay index between 42 and 105 Mm. 
This may be because of the limitations to our decay index estimates 
described in the previous section or because our sample is too 
small to detect a slight difference. 
Interestingly, \citet{Liu08} finds there to be no significant difference 
between the average decay indexes of kink and torus instability 
events, in contrast to the MHD simulations, and   
limitations to the PFSS modeling are also cited as a 
possible reason for this discrepancy. 
The fact that we find no difference in the decay indexes between events 
with and without apparent twisting motions therefore does not 
preclude differences in their initiation mechanisms. Moreover, we 
do not expect that all events in the twisted sample are  
initiated by the kink instability--most probably are not.  
Rather, we suggest simply that the likelihood is 
greater than for events without apparent twisting motions, which might explain 
the difference in onset height distributions. 
It may also be that more highly twisted filaments are likelier to facilitate a reconnection 
process at some point during their slow-rise phases before the torus instability threshold is reached. 
 
The velocities plotted in the three rightmost panels of Figure~\ref{f-twist_dist} also 
tell an interesting story. We see from the middle panel that events 
with and without apparent twisting motions have no general differences in their 
slow-rise velocities. 
If, as we hypothesize, the twisting motions indicate greater 
helicity on average, then this similarity suggests that the flux rope helicity upon eruption 
is largely decoupled from the slow-rise velocity. This constraint is  
useful for modeling efforts, and it might lead 
one to speculate that there is a comparative homogeneity 
in the slow-rise mechanisms across events with disparate 
fast-rise mechanisms.  
As with the onset heights, velocity and acceleration differences emerge between 
the twist groups during the fast rise phase. 
We find that events with apparent twisting motions compared 
to those without have final speeds in our kinematics study that 
are 1.3$\times$ faster and CME speeds across the entire 
catalog that are 1.2$\times$ faster on average. While these 
enhancements are not very large, the cumulative distributions 
in Figure~\ref{f-twist_dist} indicate that the differences are statistically 
significant, with $A^2_k$ = 1.9 \& $p$ = 0.05 for the AIA speeds 
and $A^2_k$ = 3.4 \& $p$ = 0.01 for the CME speeds. 
This suggests   
that flux rope helicity may be positively correlated with CME speed, 
though a quantitative measure of helicity is needed to make this claim. 

Two previous studies are particularly relevant to this possibility. 
\citet{Sung09} found a positive correlation between the square of 
the initial CME speed and the magnetic helicity per unit length 
of the corresponding magnetic clouds at 1 AU for 34 events. 
\citet{Park12} examined the average helicity injection ($\dot{H}$) of 
CME-producing active regions for a few days leading up to 47 eruptions 
using Michelson Doppler Imager (MDI) magnetograms and found a 
positive correlation between $\dot{H}$ and CME speed. 
This result is consistent with our finding, though it relates specifically to  
active region CMEs, while 
we find roughly the same enhancement across all filament types. 
However, when divided by type, our result remains statistically significant 
only for QS and PC events ($A^2_k$ = 3.2 \& $p$ = 0.02), for which there 
are over twice as many events compared the AR and IP sample ($A^2_k$ = -0.12 \& $p$ = 0.39).
We 
also note from \S\ref{twist} that confined eruptions are significantly 
less likely to exhibit twisting motions (44\% vs. \AllTwiYp{} for the general population), 
implying lesser helicity on average. 
Confined events are often arrested by the overlying 
field because there is insufficient energy released 
to push through, and the fact that we find significantly 
lower rates of apparent twisting motions in these eruptions 
suggests, albeit somewhat tenuously, that we are 
reasonable in considering the twisted sample to be 
generally more energetic. 

\subsection{Future Work}
\label{future}

A recently-submitted paper by \citet{Su15} presents detailed modeling and 
observations for one of the polar crown events in our sample, which  
includes a more careful estimate of the decay index 
that employs STEREO observations to better constrain the location 
of the filament. 
This study implicates the torus instability as the driving mechanism through 
a comparison of the decay index to our fast-rise onset height. 
\citet{Reeves15} present a related study on a prominence eruption 
observed by the Interface Region Imaging Spectrograph (IRIS) that 
includes a comparison of the eruption kinematics  
to brightenings observed by both IRIS and AIA, which suggest 
that reconnection below the prominence triggered the fast-rise phase. 
The techniques employed by these works can be readily applied to our 
kinematics sample to estimate relative frequencies of the 
fast-rise onset mechanisms. 
The suggestive relationship between 
apparent twisting motions and CME speed can also be 
explored by quantifying the twist or overall helicity in 
quiescent filaments. 
This might be done using helicity injection 
measurements similar to that of \citet{Park12} applied to HMI observations of events 
in the catalog observed on the disk or west limb, though the low flux 
densities in quiet sun regions may preclude reliable measurements.
Alternatively, observations of untwisting motions 
similar to those presented by \citet{Yan14} might be used to 
measure the extent to which filaments are twisted 
and then to compare with CME speed measurements.


 
\section{Conclusion} %
\label{conclusion} %

This paper details the first statistical study of prominence eruptions 
observed by the Atmospheric Imaging Assembly (AIA) aboard the 
Solar Dynamics Observatory (SDO).
An online catalog containing \TypeAll{} eruptions was developed, 
listing several properties for each event that are intended to 
aid the community in identifying eruptions for future work (\S\ref{observations}). 
Filaments are typed based on environment (active region (AR), 
intermediate (IP), quiescent (QS), or polar crown (PC)) along with 
the extent to which they erupt (full, partial, or confined), 
and comparisons between the type groups are made throughout (\S\ref{type}). 
A sample of 106 limb events was drawn from the catalog for a 
kinematics study that characterizes the distinct slow- and 
fast-rise phases often exhibited by prominence eruptions (\S\ref{kinematics}).
Our main results are summarized below: 

\begin{enumerate}

\item{
Symmetric eruptions are found to be somewhat more common than 
asymmetric (\AllSymSp{} vs. \AllSymAp{}), and symmetric events 
are 1.5$\times$ more likely to be full eruptions 
due to enhanced mass draining and detachments in asymmetric events (\S\ref{symmetry}).
Relatively few eruptions exhibit non-radial (\AllDirNRp{}) and sideways (\AllDirSp{}) 
trajectories.
Sideways trajectories are more commonly found in 
AR and IP events, which are likelier to be deflected strongly away from ARs. 
Sideways events are also more frequently confined or partial eruptions, likely 
because of interactions with neighboring structures that may 
arrest the eruption or facilitate additional draining (\S\ref{direction}).
}

\item{
Vertical threads, perpendicular to the filament spines, are observed for 
\AllVerYp{} of events, which is interesting given their perplexing nature. 
The fraction of events with vertical threads by 
type scales with the characteristic sizes of the four types, with \ARVerYp{} 
for AR and \PCVerYp{} for PC eruptions. 
We note that vertical threads appear in ARs only for particularly large 
filaments, often in decaying active regions with well-separate polarities 
or at the boundary of what might instead be 
considered an intermediate filament (\S\ref{threads}).
Coronal cavities are also found most frequently in PC events, as expected.
Overall, cavities are absent in \AllCavNp{} 
of limb events, present in \AllCavYp{}, and become visible 
during the eruptions in \AllCavDp{} (\S\ref{cavity}).
}

\item{
\AllCMEYp{} of our filament eruptions are associated with white-light 
CMEs observed by LASCO, and similar rates are found for each of the filament 
types. 
The CME speeds associated with AR and IP events are generally faster 
than those associated with QS and PC events by a factor of 1.3. 
IP-CMEs tend to be 1.1$\times$ faster than AR-CMEs, but this difference 
is not statistically significant (\S\ref{cmes}).
}

\item{
106 limb events were selected from the catalog for a kinematics study, 
including 59 QS, 31 PC, 11 IP, and 6 AR prominence eruptions. 
Height-time plots were constructed for each event using 304 \AA{} observations, 
and the Canny edge detection algorithm was used to extract individual 
height measurements (\S\ref{procedure}). 
A majority (74\%) of the rise profiles are well-fit by an 
approximation introduced by \citet{Cheng13}, which combines a linear 
equation to treat the slow-rise phase and an exponential to treat the fast-rise. 
The onset of the fast-rise phase is defined as the point at 
which the exponential component of the curve equals the linear (\S\ref{equation}). 
We also illustrate that apparent rise 
and fall velocities approaching 1 \kms{} can be observed due to solar 
rotation alone (\S\ref{rotation}).
}

\item{
The average values for the radial onset height ($h_{\rm onset}$), 
relative onset height ($\Delta{}h_{\rm{onset}}$), and slow-rise ($\Delta{}t_{\rm{slow-rise}}$) 
phase duration are 83 Mm, 29 Mm, and 4.4 hours, respectively. 
As expected, the lowest heights and shortest slow-rise phases are found 
for AR eruptions, and the largest heights and longest slow-rise phases 
are found for PC eruptions (\S\ref{onset}).
The average values for the slow-rise velocity ($v_{\rm{slow-rise}}$), 
maximum FOV velocity ($v_{\rm{fov-max}}$), and 
velocity at 500$''$ above the limb ($v_{\rm{500''}}$) are 2.1, 106, and 183 \kms{}, respectively. 
The average 
onset acceleration ($a_{\rm{onset}}$), maximum FOV acceleration ($a_{\rm{fov-max}}$), 
and acceleration at 500$''$ ($a_{\rm{500''}}$) are   
1.9, 111, and 199 \msq{}, respectively. 
As with CME speed, we find that IP events are generally the fastest of the 
four types (\S\ref{velocity}).
See Table~\ref{t-lognormal} 
for the full list of statistical parameters and Table~\ref{t-kin_types} for type 
comparisons.
}

\item{
The kinematics parameter distributions are best described by 
lognormal probability distributions similar to that of CME speeds, 
indicating that this pattern emerges in the low-corona (\S\ref{results}).
An exception to this is the slow-rise phase velocity distribution 
for PC and possibly AR eruptions, which follow Gaussian 
distributions that shift to log-normality during the fast-rise phase. 
This may suggest a  
uniformity in the slow-rise process for PC and possibly AR 
filament eruptions compared to their QS and IP counterparts, 
whose more varied environments and size scales may provide more 
diverse influences (\S\ref{velocity}).
}

\item{
We find a positive correlation between latitude and fast-rise 
onset height, 
with correlation coefficients ($r$) of 0.30 and 0.43 for the 
radial and relative onset heights, respectively. 
A corresponding correlation between latitude and slow-rise 
phase duration is also found ($r$ = 0.50).
This is a global effect that diminishes if the eruptions are 
binned into latitude groups and is therefore likely to be a product of the global 
magnetic field. 
We interpret these correlations in terms of the decay index of the vertical magnetic field 
strength, which we find to exhibit a negative correlation with latitude 
using average decay indexes obtained from 
PFSS extrapolations between heights of 42 and 105 Mm. 
High-latitude events thus tend to have larger onset heights and longer 
slow-rise phases because they tend to have smaller decay indexes at a given height. 
We also find that the distribution of decay indexes at the onset 
of the fast-rise phase spans the full range of theoretical 
torus instability or loss of confinement critical thresholds ($\sim$1--2), with an average 
value (1.1) consistent with the straight current channel approximation. 
A number of events also exhibit critical decay indexes of $<1.0$, 
indicating initiation by other means, such as reconnection or 
the kink instability (\S\ref{latitude}). 
}

\item{
A majority of events (\AllTwiYp{}) exhibit apparent twisting motions, 
and IP filaments have the highest rate (\IPTwiYp{}).
This includes several types of twist, of which 
untwisting motions seen in 
filament legs in the latter stages of an eruption are most common. 
We hypothesize that events with twisting motions 
have greater helicity on average, which is important to our interpretation of 
the relationships found between twist and eruption kinematics (see below). 
Confined eruptions exhibit twisting motions at a much 
lower rate (44\%), which we interpret as a consequence of their comparatively limited
magnetic energy (\S\ref{twist}).
A small fraction of events exhibit writhing motions suggestive 
of the kink instability (\AllKinYp{}), but 
half of these are labeled as ambiguous. As expected, writhed events are much more likely 
to be partial or confined than the general population (\S\ref{kink}).
}

\item{
Statistically significant differences are found between 
events that exhibit signs of twist and those that 
do not. 
Twisted events transition into their fast-rise phases at 
lower heights and have faster low-corona speeds, which 
might reflect different populations of  
fast-rise onset mechanisms that preferentially affect 
more or less twisted flux ropes. 
We also find that events with apparent 
twisting motions have faster 
coronagraph CME speeds by a factor of 1.2 across all 
events in the catalog. Though 
not very large, this difference is statistically 
significant ($p$ = 0.01).
\citet{Park12} showed that helicity injection is positively correlated 
with CME speed for AR events, and  
our results indicate the same relationship for filament eruptions of all types. 
We also find that there is no difference in slow-rise 
speed for events with and without apparent twisting motions, which 
may imply that the flux rope helicity upon eruption 
is largely decoupled from the slow-rise mechanism 
and that there may be a comparative homogeneity 
in the slow-rise mechanisms across events with disparate fast-rise mechanisms (\S\ref{speed_twist}).
}

\end{enumerate}

\begin{acks}
Support for this work was provided by the National Aeronautics and Space Administration (NASA) 
through grant NNX12AI30G to 
the Smithsonian Astrophysical Observatory (SAO), 
by the National Science Foundation (NSF) through grant AGS1263241 for the solar 
physics Research Experiences for Undergraduates (REU) program at SAO, and  
by the Lockheed-Martin Solar and Astrophysics Laboratory (LMSAL) through contract 
SP02H1701R to SAO for support of the AIA. 
Additional support was provided by the National Science Foundation of China (NSFC) 
through grants No. 11333009, 11173062, 11473071, and J1210039, along with the 
Youth Fund of Jiangsu through grant No. BK20141043.
The SDO is a NASA satellite, and the 
AIA instrument team is led by LMSAL. 
We gratefully acknowledge the anonymous referee for their constructive comments. 
P.I.M. thanks Sarah Gibson for her FORWARD tutorial, which 
facilitated our decay index analyses. 
We also thank the observers 
who contributed filament eruptions to the HEK: Anna Malanushenko, Nariaki Nitta, 
Wei Liu, Karel Schrijver, Mark Cheung, Ryan Timmons, Thomas Berger, Marc Derosa, 
Ralph Seguin, Paul Higgins, Juan Mart\'{i}nez-Skyora, Alberto Sainz-Dalda, Gregory Slater, 
and Neil Hurlburt. 
\end{acks}


  
\bibliographystyle{spr-mp-sola}

\tracingmacros=2
\bibliography{filament_catalog_references}  

\IfFileExists{\jobname.bbl}{} {\typeout{}
\typeout{****************************************************}
\typeout{****************************************************}
\typeout{** Please run "bibtex \jobname" to obtain} \typeout{**
the bibliography and then re-run LaTeX} \typeout{** twice to fix
the references !}
\typeout{****************************************************}
\typeout{****************************************************}
\typeout{}}

\end{article} 

\end{document}